\PassOptionsToPackage{numbers,sort&compress}{natbib}
\documentclass{article}

\usepackage[preprint]{neurips_2026}

\usepackage[utf8]{inputenc}
\usepackage[T1]{fontenc}
\usepackage{hyperref}
\usepackage{url}
\usepackage{booktabs}
\usepackage{amsfonts}
\usepackage{amsmath}
\usepackage{amssymb}
\usepackage{algorithm}
\usepackage{algpseudocode}
\usepackage{nicefrac}
\usepackage{microtype}
\usepackage{xcolor}
\usepackage{graphicx}
\usepackage{multirow}
\usepackage{subcaption}
\usepackage{booktabs}
\usepackage[normalem]{ulem}
\setcitestyle{numbers,square,comma}

\definecolor{americanrose}{rgb}{1.0, 0.01, 0.24}

\title{QuLoC: Photonic \textbf{Qu}antum-Assisted \textbf{Lo}w-Rank LLM \textbf{C}ompression}

\author{%
  Xiao-Hui Ni$^{1}$\thanks{These authors contributed equally
  to this work and share first authorship.}
  \quad
  Yu-Han Yao$^{1,2}$\footnotemark[1]
  \quad
  Yu-Ze Zhu$^{1}$
  \quad
  Hang Song$^{3}$ \\[0.3em]
  \textbf{Xiang Zhao}$^{1}$\thanks{Corresponding authors:
  \texttt{zhaoxiang@turingq.com},
  \texttt{yanglin@turingq.com},
  \texttt{xianmin.jin@sjtu.edu.cn}.}
  \quad
  \textbf{Lin Yang}$^{1}$\footnotemark[2]
  \quad
  \textbf{Xian-Min Jin}$^{1,3,4,5,6}$\footnotemark[2]
  \\[0.6em]
  $^{1}$TuringQ Co., Ltd., Shanghai 200240, China
  \\[0.3em]
  $^{2}$Department of Information and Communication Engineering, \\
  Graduate School of Information Science and Technology, \\
  The University of Tokyo, Tokyo 113-8656, Japan
  \\[0.3em]
  $^{3}$Center for Integrated Quantum Information Technologies (IQIT), \\
  School of Physics and Astronomy and State Key Laboratory of \\
  Photonics and Communications, \\
  Shanghai Jiao Tong University, Shanghai 200240, China
  \\[0.3em]
  $^{4}$Chip Hub for Integrated Photonics Xplore (CHIPX), \\
  Shanghai Jiao Tong University, Wuxi 214000, China
  \\[0.3em]
  $^{5}$Atomology Co., Ltd., Shanghai 200240, China
  \\[0.3em]
  $^{6}$Hefei National Laboratory, Hefei 230088, China
}

\begin{document}

\maketitle

\begin{abstract}
As LLMs grow in size, compression becomes increasingly important for efficient deployment.
SVD-based low-rank compression reduces parameter counts but can degrade downstream performance.
To improve performance after compression, we introduce QuLoC, a photonic quantum-assisted LLM compression algorithm that uses quantum circuit outputs to gate the retained low-rank components during training.
Model performance is recovered through local functional reconstruction followed by end-to-end knowledge distillation.
After training, the gating coefficients are absorbed into the low-rank factors, allowing the compressed model to run on classical hardware without executing quantum circuits during inference.
Experiments on Qwen3.5-4B show that QuLoC achieves a 9.73\% relative improvement in average accuracy over state-of-the-art baselines across multiple downstream tasks, demonstrating its effectiveness.
We further evaluate QuLoC on LLaMA-7B at different parameter compression ratios.
It achieves comparable or higher average downstream accuracy than state-of-the-art baselines, supporting its applicability to a larger model across different compression settings. 
These results motivate further exploration of photonic quantum-assisted compression for larger models and more complex agentic tasks.
\end{abstract}

\section{Introduction}
\label{sec:intro}

Large language models (LLMs) have achieved substantial improvements in language understanding and generation through increases in model size, training data, and computational resources~\cite{LLM_overview2025,zhao2026survey,chang2024survey,zhang2025survey}.
However, their large parameter counts lead to substantial model storage and inference memory requirements, making deployment costly, particularly in resource-constrained environments~\cite{fernandez-etal-2025-energy,LLM-inference}.
Therefore, reducing the number of model parameters while preserving language modeling and downstream task performance is an important goal of LLM compression~\cite{AWQ,Losparse,PV-Tuning}.

To this end, several approaches have been explored, including structured pruning~\cite{ma2023llm,qu2025automatic,hedegaard2024structured,ICLR2024_160adf2d,li2026sepprune,an2024fluctuation}, knowledge distillation into smaller student models~\cite{yang2025survey,fang2025knowledge,du2025active,wang2026end,li2026backdoor}, and low-rank decomposition~\cite{Dobi-SVD,GF-SVD,QSVD,saha2024llm,zhou2026llm}.
Structured pruning reduces model size by removing structured groups of parameters, such as those associated with selected channels or attention heads~\cite{zhang2024loraprune}. For model compression, knowledge distillation typically transfers knowledge from a larger teacher to a smaller student by training the student with teacher-provided supervision~\cite{gu2024minillm}. A common approach to low-rank compression approximates a large weight matrix as the product of two smaller matrices whose shared output dimension is specified by a rank parameter~\cite{wang2025svdllm}. Choosing a sufficiently small rank reduces the combined parameter count, while further rank reduction restricts the representational capacity of the approximation and may compromise model performance.

For a given rank, truncated singular value decomposition (SVD) constructs a low-rank approximation by retaining the leading singular components~\cite{wang2025svdllm,wang2025svdllmv2}. Although this approximation minimizes weight reconstruction error under the Frobenius norm, it does not necessarily preserve the model's behavior on actual inputs. Previous studies have shown that optimizing low-rank factors after decomposition can help recover performance lost through compression~\cite{sy2025efficient,ren2023lowrank}. Related work on low-rank adaptation uses explicit gating vectors or trainable diagonal coefficients to control which components are retained and how parameter budgets are allocated~\cite{ding2023sparse,zhang2023adalora}. Together, these studies motivate us to introduce explicit channel-wise gating into trainable low-rank compression and investigate whether jointly optimizing the gating mechanism and the low-rank factors can further improve the compressed model's performance at a given rank.

Parameterized photonic quantum circuits offer a possible route to generating these gating coefficients. Specifically, their photon detection probabilities vary with the circuit parameters and can be mapped to channel-wise scaling coefficients through a compact classical mapping. This allows the circuit parameters to be optimized jointly with the low-rank factors. Whether this photonic quantum-assisted approach to low-rank LLM compression can achieve performance comparable to or better than classical compression methods remains an open research question.

To investigate this question, we propose a photonic \textbf{Qu}antum-assisted \textbf{Lo}w-rank LLM \textbf{C}ompression (QuLoC) algorithm that uses photonic quantum circuit outputs to generate gating coefficients, which explicitly scale the contributions of individual low-rank channels during training.
It then further recovers model performance through two-stage training that combines local functional reconstruction with end-to-end knowledge distillation. After training, the gating coefficients are absorbed into the low-rank factors, allowing the compressed model to run on classical hardware without executing quantum circuits during inference.

By using photonic quantum circuit with sixteen modes, experiments on Qwen3.5-4B show that QuLoC achieves a 9.73\% relative improvement in average accuracy across six downstream tasks over the strong classical baseline Swift-SVD~\cite{swift-svd}, while reducing the language-model parameter count by 20.05\%.
These results demonstrate the effectiveness of QuLoC in preserving downstream performance at a higher parameter reduction ratio.
We further evaluate QuLoC on LLaMA-7B at model parameter retention ratios of 80\% and 60\%. Its mean downstream accuracies are 52\% and 44\%, respectively. Compared with Swift-SVD*, QuLoC is one percentage point higher at 80\% retention and ties at 60\% retention at the reported precision. Additionally, we evaluate QuLoC using photonic quantum circuits with four modes and a squeezed vacuum state of light.
Its average downstream accuracy is only 1.48\% lower than that of the sixteen-mode implementation, while remaining higher than that of the baselines.

In summary, the main contributions of this paper are summarized as follows:
\begin{itemize}
    \item We propose a photonic quantum-assisted low-rank compression algorithm. The low-rank factors reduce the parameter count of the original projection matrices, while the gating vector generated from photonic quantum circuit outputs further adjusts the contribution of each rank channel during training without restoring the original full-size weight matrices.
    
    \item We introduce a two-stage training strategy consisting of local functional reconstruction followed by end-to-end knowledge distillation. The first stage preserves the local input–output behavior of the replaced modules, while the second jointly optimizes the compressed modules within the full student model
    to recover overall language modeling capability and account for interactions among compressed layers.
    
    \item We fold the optimized gating vector into the low-rank factors, thus eliminating the need for photonic quantum hardware during deployment.
    
\end{itemize}

The remainder of this paper is organized as follows. Section~\ref{sec:related} reviews related work on large language model compression. Section~\ref{sec:method} presents the overall QuLoC framework, including local reconstruction, the two-stage  complementary training strategy, and the classical deployment procedure. Section~\ref{sec:experiments} and Section~\ref{sec:analysis} present the experimental setup and results, respectively. Finally, Section~\ref{sec:discussion} discusses the experimental findings, interprets the roles of the proposed components, and analyzes the limitations of the framework, while Section~\ref{sec:conclusion} summarizes the main findings and outlines future research directions.

\section{Related Work}
\label{sec:related}

This section reviews two lines of research relevant to QuLoC: classical low-rank compression of large language models, and quantum-assisted model compression. The first subsection reviews representative classical low-rank compression methods that aim to reduce model parameters while preserving model performance. The second subsection reviews representative quantum-assisted model compression methods aimed at reducing model parameter counts. Through this review, we aim to help readers understand how QuLoC differs from existing approaches.

\subsection{Classical Low-Rank Compression of Large Language Models}
\label{sec:related_low_rank_mlp}

Low-rank compression approximates dense weight matrices as products of two lower-dimensional factors, and it can significantly reduce the parameter count when the rank is sufficiently small. Truncated singular value decomposition (SVD) provides an optimal rank-$r$ approximation of a weight matrix under the Frobenius norm, but minimizing weight reconstruction error does not necessarily preserve the model's behavior on actual inputs. To better preserve model performance, existing methods use parameter sensitivity or input activation statistics to guide low-rank approximation. Specifically, Fisher-weighted SVD (FWSVD) weights reconstruction errors according to Fisher information, placing greater emphasis on preserving parameters that are more sensitive to the task loss~\cite{FWSVD}. Activation-aware SVD (ASVD) instead uses input activation statistics to rescale weight matrices before decomposition, accounting for the unequal contributions of input channels to layer outputs~\cite{ASVD}.

However, incorporating parameter importance or activation statistics does not by itself guarantee that the singular values of the transformed weights directly reflect their contributions to layer output reconstruction error. Consequently, discarding the smallest singular values may not minimize this error. To address this mismatch, SVD-LLM introduces truncation-aware data whitening, which establishes a direct relationship between the discarded singular values and the resulting output reconstruction loss. It further updates the low-rank factors sequentially to recover performance after truncation~\cite{wang2025svdllm}.

Beyond improving the decomposition of individual matrices, allocating the compression budget across matrices is also important because their redundancy and sensitivity to truncation differ. SVD-LLM V2 addresses this heterogeneity by assigning matrix-specific compression ratios based on theoretical truncation loss and introducing loss-optimized weight truncation to reduce discrepancies between theoretical and practical reconstruction losses~\cite{wang2025svdllmv2}. In parallel, Dobi-SVD approaches compression through activation truncation, using differentiable optimization to select truncation positions and incremental principal component analysis to reconstruct the corresponding weights~\cite{Dobi-SVD}.

Despite these advances, efficiently computing activation-aware low-rank approximations and properly allocating ranks to preserve overall model performance remain important challenges. More recently, Swift-SVD addresses the computational challenge by incrementally accumulating output activation covariance and extracting its dominant directions through a single eigendecomposition. This construction minimizes layer-wise output reconstruction error at a given rank and supports different rank choices without repeating the decomposition. To better preserve overall model performance, Swift-SVD allocates ranks using both local reconstruction loss and layer importance, reserves a minimum rank for each matrix, and selects among candidate allocations based on validation performance. This strategy balances local approximation quality with layer importance while limiting excessive compression of individual matrices~\cite{swift-svd}.

These studies establish effective classical approaches to constructing and refining low-rank representations. Building on this foundation, QuLoC investigates photonic quantum-assisted modulation of low-rank channels during training. Rather than generating the full low-rank factors from quantum outputs, it uses these outputs to determine channel-wise gating coefficients and jointly optimizes the resulting representation through local functional reconstruction and end-to-end knowledge distillation.

\subsection{Quantum-Assisted Model Compression}
\label{sec:related_quantum_assisted_compression}

Knowledge distillation transfers knowledge from a larger teacher to a smaller student, offering a means of reducing model parameters and storage requirements. Quantum-assisted variants either use quantum circuits to replace all or part of the classical student network or incorporate quantum-information-based operations into the distillation objective or student parameterization~\cite{nishan2026quantummedkd,li2026quantumgated,krishnamurthy2026early,alam2023approximate}. Existing approaches relevant to this work can be broadly grouped into two categories based on where quantum modules are incorporated into the distillation process. The first distills knowledge from a classical teacher into a quantum student, while the second uses quantum-assisted training to distill knowledge from a classical teacher into a classical student.

\paragraph{Classical-to-quantum distillation.}

In this approach, knowledge is transferred from a classical neural network to a quantum or hybrid student containing a parameterized quantum circuit. In this route, the classical teacher provides predicted class probabilities as supervisory signals for the quantum student. The student is trained by optimizing its quantum circuit parameters and any trainable classical parameters using a distillation loss that matches the teacher's predicted probabilities and a supervised classification loss based on ground-truth labels~\cite{hasan2023bridging,li2025quantumllm}. Existing studies have explored both broader application settings and different forms of distillation supervision. 

In representative studies, Hasan and Mahdy use classical CNN predictions to supervise 4- and 8-qubit quantum students for image classification. On MNIST, distillation from AlexNet improves the accuracy of the 4-qubit student from 77.69\% to 85.18\%, a gain of 7.49 percentage points over training without distillation, thus demonstrating that teacher supervision can improve the predictive performance of quantum models with limited qubit resources~\cite{hasan2023bridging}. QD-LLM extends this approach to classical LLM teachers and quantum students for text classification, combining KL and Jensen--Shannon divergences for teacher-student prediction matching with cross-entropy supervision from ground-truth labels~\cite{li2025quantumllm}. Beyond output distributions, Barbato et al.\ transfer intermediate representations to a hybrid quantum-classical student, using a regression module to match feature dimensions and an MSE alignment loss alongside supervised classification~\cite{barbato2026hybridcompression}. These studies extend classical-to-quantum distillation from image to text classification and from prediction matching to intermediate feature alignment.

In these approaches, the quantum circuit remains part of the student's inference computation. Deployment therefore requires quantum circuit , either on quantum hardware or through classical simulation. Moreover, reducing the number of trainable parameters does not necessarily reduce deployment cost. Specifically, the latter also depends on classical pre-processing and network components, circuit width and depth, measurement requirements, and the simulation or hardware implementation. In addition, the cited methods primarily transfer predictive behavior or intermediate representations into a quantum or hybrid student. They do not directly provide a factorization of the high-dimensional classical projection matrices within a Transformer-based language model, such as its MLP weight matrices.

\paragraph{Quantum-assisted classical distillation.}
To avoid the dependence of inference on quantum hardware, another line of research uses quantum circuits or quantum-information techniques only as auxiliary modules during training, while both the teacher and the student remain classical neural networks~\cite{liu2026qrkd,chen2026photonic}. In this route, quantum modules can be used to construct distillation objectives, transform intermediate features, or generate auxiliary signals for the parameterization of the classical student model. After training, the resulting student can perform inference on classical computing hardware.

Quantum Relational Knowledge Distillation (QRKD) maps classical hidden features into a quantum Hilbert space and uses quantum kernels to characterize inter-sample relationships. Distillation then aligns the relational information associated with the teacher and student, rather than relying exclusively on pointwise prediction matching~\cite{liu2026qrkd}. QRKD has been evaluated on GPT-2 language-modeling benchmarks, including WikiText-2 and Penn Treebank. Since the quantum kernels are used only to compute the training objective, the trained classical student requires no quantum computation during inference. 

Unlike QRKD, Photonic Quantum-Enhanced Knowledge Distillation (PQKD) introduces a structured parameterization of CNN student weights~\cite{chen2026photonic}. Each convolutional kernel is represented using trainable spatial basis filters and channel-mixing coefficients generated from photonic measurement features through a fixed linear map. This construction reduces the number of independently trainable parameters. Training alternates between gradient-based student updates and sampling-based photonic-parameter updates. Under its fixed-input formulation, the generated coefficients could be cached after training to permit classical inference. Experiments on image classification examine the trade-off between trainable parameter reduction and predictive accuracy. These reductions do not directly establish equivalent savings in deployment memory or computation.

In summary, QRKD and PQKD introduce quantum assistance at two different technical levels. Specifically, QRKD incorporates quantum relational information into the distillation objective to improve the transfer of inter-sample relationships from teacher to student, whereas PQKD uses photonic circuit outputs to parameterize the classical student's convolutional weights, reducing the number of independently trainable parameters. Inspired by PQKD, QuLoC introduces trainable photonic gating coefficients to modulate the relative contributions of SVD-initialized low-rank components, thus addressing the limited adaptability caused by fixed component weights during performance recovery.

\section{Proposed Method}
\label{sec:method}

This section presents the QuLoC method.
We first outline the overall workflow in Section~\ref{sec:method_overview}, then introduce the photonic gating mechanism and the low-rank parameterization of SwiGLU projections
in Section~\ref{sec:photonic_low_rank}.
To recover performance after compression, we describe Stage~I
local functional reconstruction in
Section~\ref{sec:local_reconstruction}, followed by Stage~II
end-to-end knowledge distillation in
Section~\ref{sec:end_to_end_distillation}.
Once training is complete, Section~\ref{sec:gain_folding}
explains how the gating coefficients are folded into the low-rank
factors for classical inference.
Finally, Section~\ref{sec:complexity} analyzes the total model parameter counts.

\subsection{Method Overview}
\label{sec:method_overview}

\begin{figure}
    \centering
    \includegraphics[width=\linewidth]{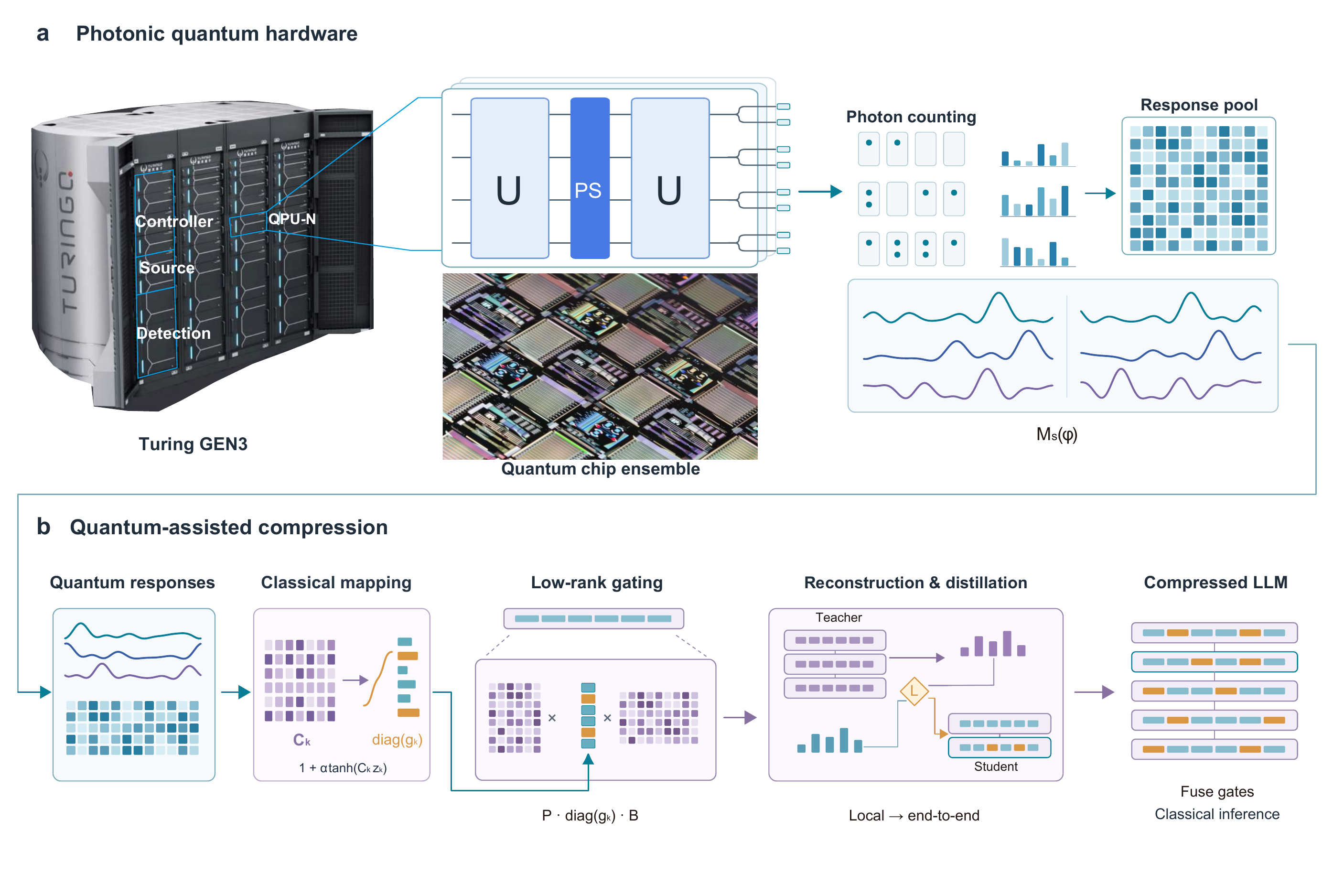}
    \caption{Photonic quantum-assisted low-rank compression of large language models. (a) Photon-counting measurements from an ensemble of four-mode photonic quantum circuits are used to build a reusable pool of quantum response features. (b) Lightweight classical mappings transform these features into input-independent gates that modulate the low-rank projection factors during training. The compressed student is optimized through layer-wise reconstruction followed by end-to-end teacher–student distillation. After training, the gates are folded into the low-rank factors, enabling inference entirely on classical hardware.}
    \label{fig:placeholder}
\end{figure}

This section presents QuLoC, a photonic quantum-assisted compression method that combines parameterized photonic circuits with a two-stage training strategy to recover performance after compression. Figure~\ref{fig:placeholder}  illustrates how photonic response features are obtained from four-mode circuits and incorporated into the QuLoC compression workflow.

Considering that MLP blocks account for a substantial fraction of the model parameters and contain large dense projection matrices suitable for low-rank factorization, we select them as the targets of low-rank compression.
Specifically, the 32 SwiGLU MLP blocks in Qwen3.5-4B contain approximately 2.265 billion parameters. Low-rank compression of these projections can substantially reduce the model's parameter count while leaving the attention modules and other model components unchanged. The necessary preliminaries on Transformer-based MLP architectures and photonic quantum circuits are provided
in Appendix~\ref{sec:preliminary}.

For each selected SwiGLU MLP, QuLoC replaces the gate,
up, and down projections with SVD-initialized low-rank
factors.
Each projection is associated with a separate trainable
photonic quantum circuit.
By a compact classical mapping matrix, the circuit's photon detection probabilities are mapped
to a gain vector, whose entries are multiplicative
coefficients that scale the corresponding low-rank
bottleneck activations.
We refer to this mechanism as \emph{photonic gating}. 

In Stage~I, each compressed MLP is trained independently to minimize the discrepancy between its output and that of its corresponding uncompressed MLP. Notably, each compressed MLP receives the same input as its corresponding uncompressed MLP during local reconstruction. Then, the reconstructed modules are assembled into a student model. However, errors from earlier compressed modules can alter the inputs to later modules and propagate through the model, so accurate local reconstruction does not necessarily ensure good overall performance. To address this mismatch between local reconstruction and whole-model behavior, QuLoC introduces Stage~II, where end-to-end knowledge distillation jointly optimizes the compressed modules to mitigate these effects and improve the student model's performance.

After the end-to-end training, the final gain vectors can be folded exactly into the low-rank factors because they are independent of the MLP input. Therefore, the deployed model performs inference entirely using classical operations, without requiring photonic quantum circuit evaluation.

\subsection{Photonic Gating for Low-Rank SwiGLU}
\label{sec:photonic_low_rank}

For each SwiGLU MLP selected for compression, we represent
its gate, up, and down projection matrices using low-rank
factors $P_k$ and $B_k$ with photonic gating:
\begin{equation}
W_k \approx \widetilde{W}_k
= P_k \operatorname{diag}(\mathbf g_k) B_k,
\label{eq:photonic_low_rank}
\end{equation}
where
$k\in\{\mathrm g,\mathrm u,\mathrm d\}$ indexes the gate,
up, and down projections, respectively. Here, for
$W_k\in\mathbb R^{{m_k}\times {d_k}}$,
the trainable low-rank factor
$B_k\in\mathbb R^{r_k\times d_k}$ maps the input to an $r_k$-dimensional representation.
The gating vector $\mathbf g_k\in\mathbb R^{r_k}$
scales the individual channels of this representation
through $\operatorname{diag}(\mathbf g_k)$, and
$P_k\in\mathbb R^{m_k\times {r_k}}$
maps the scaled representation to the output space.
Here, $d$ and $m$ are input and output dimensions of the projection matrix, respectively. 
The gating vector is generated by mapping the output
probabilities of a photonic quantum circuit through a
trainable classical matrix. Below, we describe the
initialization of $P_k$ and $B_k$ and the generation
of $\mathbf g_k$.

\paragraph{The initialization of $P_k$ and $B_k$.} For each projection matrix $W_k$, we initialize its
low-rank factors using truncated singular value
decomposition:
\begin{equation}
\begin{aligned}
W_k
&\approx U_{k,r_k}\Sigma_{k,r_k}V_{k,r_k}^{\top} \\
&=\left(U_{k,r_k}\Sigma_{k,r_k}^{1/2}\right)
  \left(\Sigma_{k,r_k}^{1/2}V_{k,r_k}^{\top}\right) \\
&=P_kB_k,
\end{aligned}
\label{eq:low_rank_factorization}
\end{equation}
where $r_k$ is the retained rank of projection $k$.
We set $r_{\mathrm g}=r_{\mathrm u}=r$ and
$r_{\mathrm d}=2r$. Since the down projection maps the intermediate features back to the residual stream and directly determines the MLP output, we heuristically assign it a larger rank to provide greater approximation capacity.

\paragraph{The generation of the photonic gating $g_k$.} Each projection is associated with a photonic quantum circuit.
For a fixed circuit input state, let $\mathbf z_k \in\mathbb R^{d_z}$ denote a photonic feature vector obtained from photonic quantum circuits. We map this vector to a gain vector
$\mathbf g_k\in\mathbb R^{r_k}$ using a trainable
mapping matrix $C_k\in\mathbb R^{r_k\times d_{z}}$:
\begin{equation}
\mathbf g_k=\mathbf 1+\alpha\tanh(C_k\mathbf z_k),
\label{eq:channel_gain}
\end{equation}
where $\mathbf 1\in\mathbb R^{r_k}$ is the all-ones
vector and $\tanh$ is applied element-wise.
The entries of $\mathbf g_k$ multiplicatively scale
the corresponding low-rank bottleneck activations.
We restrict the hyperparameter $\alpha$ to $[0,1]$ so that the gating coefficients remain nonnegative. 
Larger values of $\alpha$ permit stronger channel
attenuation or amplification, while $\alpha=0$
disables gating. We set $\alpha=1$ in our experiments,
allowing each gain to vary within $(0,2)$.

Following the SwiGLU formulation in Appendix~\ref{sec:qwen_mlp}, we express the compressed MLP computation for an input $\mathbf x$ as
\begin{align}
\widetilde{\mathbf a}&=P_{\mathrm g}\left(\mathbf g_{\mathrm g}\odot B_{\mathrm g}\mathbf x\right), \\
\widetilde{\mathbf p}&=P_{\mathrm u}\left(\mathbf g_{\mathrm u}\odot B_{\mathrm u}\mathbf x\right), \\
\widetilde{\mathbf y}&=P_{\mathrm d}\left[\mathbf g_{\mathrm d}\odot B_{\mathrm d}\left(\operatorname{SiLU}(\widetilde{\mathbf a})\odot\widetilde{\mathbf p}\right)\right].
\label{eq:photonic_swiglu}
\end{align}
Here, $\odot$ denotes element-wise multiplication.
Unlike the gate projection in SwiGLU, photonic gating scales the low-rank bottleneck activations of all three projections.

Because photonic quantum circuits receive fixed input states rather than token-dependent inputs, the gain vector for each projection is independent of the MLP input $\mathbf x$
and is shared across tokens.
The gain vectors can change during training as the circuit parameters and mapping matrices are optimized.
After training, their final values can be folded exactly into the corresponding low-rank factors, as detailed in Section~\ref{sec:gain_folding}.

\subsection{Stage I: Layer-Wise Functional Reconstruction}
\label{sec:local_reconstruction}

Low-rank factorization changes the local mapping of the original MLP, and errors from multiple compressed MLPs may accumulate when they operate together.
We therefore reconstruct each target layer independently before end-to-end distillation, and the Stage~I reconstruction procedure is summarized in Algorithm~\ref{alg:stage1}.
For a target layer $\ell$, the uncompressed teacher performs a full forward pass and provides the layer input $\mathbf{x}_{\ell}$, the up-projection output $\mathbf{p}_{\ell}$, the fused feature $\mathbf{h}_{\ell}$, and the complete MLP output $\mathbf{y}_{\ell}$.
The compressed student module receives the same layer input $\mathbf{x}_{\ell}$ and produces the corresponding outputs $\widetilde{\mathbf{p}}_{\ell}$ and $\widetilde{\mathbf{y}}_{\ell}$.
Each target MLP is reconstructed independently using the layer input $\mathbf{x}_{\ell}$ obtained from the uncompressed teacher, rather than activations produced by a forward pass through other compressed MLPs.

For a student feature tensor $\widetilde{\mathbf F}$ and its teacher counterpart $\mathbf F$, we define the reconstruction distance over the set of valid token positions $\Omega$ as
\begin{equation}
D(\widetilde{\mathbf F},\mathbf F)=\frac{\sum_{i\in\Omega}\left\|\widetilde{\mathbf F}_{i}-\mathbf F_{i}\right\|_{2}^{2}}{\sum_{i\in\Omega}\left\|\mathbf F_{i}\right\|_{2}^{2}+\epsilon}+\beta\left(1-\frac{1}{|\Omega|}\sum_{i\in\Omega}\cos(\widetilde{\mathbf F}_{i},\mathbf F_{i})\right),
\label{eq:reconstruction_distance}
\end{equation}
where $\Omega$ is the set of valid token positions excluding padding, $i$ indexes a token position, and $|\Omega|$ denotes the number of valid positions.
$\widetilde{\mathbf F}_i$ and $\mathbf F_i$ are the student and teacher feature vectors at position $i$, respectively.
$\|\cdot\|_2$ denotes the Euclidean norm, and
$\cos(\widetilde{\mathbf F}_i,\mathbf F_i)$ denotes their cosine similarity.
The constant $\epsilon=10^{-8}$ ensures numerical stability in the denominator of the normalized squared error term, while $\beta=0.1$ controls the weight of the cosine distance term.

In addition to the complete MLP output and the up-projection output, we explicitly reconstruct the compressed down projection.
We feed the teacher fused feature $\mathbf{h}_{\ell}$ into the student down projection:
\begin{equation}
\overline{\mathbf y}_{\ell}=P_{\ell,\mathrm d}\left(\mathbf g_{\ell,\mathrm d}\odot B_{\ell,\mathrm d}\mathbf h_{\ell}\right).
\label{eq:isolated_down_output}
\end{equation}
The reconstruction terms are then defined as
\begin{equation}
\begin{aligned}
\mathcal{L}_{y}&=D(\widetilde{\mathbf{y}}_{\ell},\mathbf{y}_{\ell}), \\
\mathcal{L}_{\mathrm{up}}&=D(\widetilde{\mathbf{p}}_{\ell},\mathbf{p}_{\ell}), \\
\mathcal{L}_{\mathrm{down}}&=D(\overline{\mathbf{y}}_{\ell},\mathbf{y}_{\ell}).
\end{aligned}
\label{eq:local_reconstruction_terms}
\end{equation}
The Stage~I objective is
\begin{equation}
\mathcal{L}_{\mathrm{local}}=0.8\mathcal{L}_{y}+0.1\left(\mathcal{L}_{\mathrm{up}}+\mathcal{L}_{\mathrm{down}}\right).
\label{eq:local_objective}
\end{equation}

This objective emphasizes reconstruction of the complete MLP output through $\mathcal{L}_{y}$, with $\mathcal{L}_{\mathrm{up}}$ and $\mathcal{L}_{\mathrm{down}}$ providing auxiliary supervision for the up and down projections, respectively.

During training, we also record the gate-projection reconstruction distance
$\mathcal{L}_{\mathrm{gate}}$
and monitor it alongside $\mathcal{L}_{\mathrm{up}}$, $\mathcal{L}_{\mathrm{down}}$, and $\mathcal{L}_{y}$ to help identify potential sources of local
reconstruction error. In our experiments, $\mathcal{L}_{\mathrm{gate}}$ remained small throughout most of Stage~I training, indicating that the gate-projection outputs were
already closely aligned under the existing objective. Based on this observation, we used
$\mathcal{L}_{\mathrm{gate}}$ only as a diagnostic metric and did not include it as a separate weighted term in $\mathcal{L}_{\mathrm{local}}$. The student gate-projection parameters were still optimized through $\mathcal{L}_{y}$.

During Stage~I, we optimize the low-rank factors $P_k$ and $B_k$, the classical mapping matrices $C_k$, and the photonic-circuit parameters $\boldsymbol{\theta}_k$ for all three projections $k\in\{\mathrm g,\mathrm u,\mathrm d\}$ of each target MLP. The factors $P_k$ and $B_k$ are initialized using the truncated SVD of the corresponding original projection matrices, as defined in
Section~\ref{sec:photonic_low_rank}. The mapping matrices $C_k$ and circuit parameters
$\boldsymbol{\theta}_k$ are randomly initialized.

At each training step, we compute the gating vectors from the circuit outputs and classical mappings, evaluate $\mathcal{L}_{\mathrm{local}}$, and backpropagate through the compressed MLP and differentiable photonic circuits to obtain gradients for all trainable parameters. The classical parameters $\{P_k,B_k,C_k\}$ and quantum circuit parameters $\boldsymbol{\theta}_k$ are updated at each step using this shared objective, without separate classical and quantum training stages. A smaller learning rate is assigned to $P_k$ and $B_k$ than to $C_k$ and $\boldsymbol{\theta}_k$ to limit changes to the SVD-initialized factors. For each target layer, we retain the checkpoint
with the lowest validation $\mathcal{L}_{\mathrm{local}}$. The selected compressed MLPs are then assembled into the student model for Stage~II.

\begin{algorithm*}[h]
\caption{Stage I: Layer-Wise MLP Reconstruction}
\label{alg:stage1}
\begin{algorithmic}[1]
\Require Frozen teacher $T$; target layers $\mathcal I$;
calibration data $\mathcal D_{\mathrm{cal}}$ with a held-out validation split; base rank $r$; training steps $N_1$ \Ensure Reconstructed MLP checkpoints $\{\Theta_\ell^*\}_{\ell\in\mathcal I}$

\State Set $r_{\mathrm g}=r_{\mathrm u}=r$ and $r_{\mathrm d}=2r$
\For{each target layer $\ell\in\mathcal I$}
    \For{$k\in\{\mathrm g,\mathrm u,\mathrm d\}$}
        \State Initialize $P_{\ell,k}$ and $B_{\ell,k}$
        by rank-$r_k$ truncated SVD
        \State Fix the circuit input state
        $\lvert\psi_{\mathrm{in},\ell,k}\rangle$;
        randomly initialize $C_{\ell,k}$ and
        $\boldsymbol{\theta}_{\ell,k}$
    \EndFor
    \For{$t=1$ to $N_1$}
        \State Run $T$ on a calibration batch and obtain
        $\mathbf x_\ell$, $\mathbf p_\ell$,
        $\mathbf h_\ell$, and $\mathbf y_\ell$
        \For{$k\in\{\mathrm g,\mathrm u,\mathrm d\}$}
            \State Obtain the circuit output probabilities
            $\mathbf z_{\ell,k}$ and compute
            $\mathbf g_{\ell,k}
            \gets\mathbf 1+\alpha
            \tanh(C_{\ell,k}\mathbf z_{\ell,k})$
        \EndFor
        \State $\widetilde{\mathbf a}_\ell
        \gets P_{\ell,\mathrm g}
        \bigl(\mathbf g_{\ell,\mathrm g}
        \odot B_{\ell,\mathrm g}\mathbf x_\ell\bigr)$
        \State $\widetilde{\mathbf p}_\ell
        \gets P_{\ell,\mathrm u}
        \bigl(\mathbf g_{\ell,\mathrm u}
        \odot B_{\ell,\mathrm u}\mathbf x_\ell\bigr)$
        \State $\widetilde{\mathbf h}_\ell
        \gets \operatorname{SiLU}
        (\widetilde{\mathbf a}_\ell)
        \odot\widetilde{\mathbf p}_\ell$
        \State $\widetilde{\mathbf y}_\ell
        \gets P_{\ell,\mathrm d}
        \bigl(\mathbf g_{\ell,\mathrm d}
        \odot B_{\ell,\mathrm d}
        \widetilde{\mathbf h}_\ell\bigr)$
        \State $\overline{\mathbf y}_\ell
        \gets P_{\ell,\mathrm d}
        \bigl(\mathbf g_{\ell,\mathrm d}
        \odot B_{\ell,\mathrm d}
        \mathbf h_\ell\bigr)$
        \State $\mathcal L_{\mathrm{local}}
        \gets 0.8D(\widetilde{\mathbf y}_\ell,\mathbf y_\ell)
        +0.1D(\widetilde{\mathbf p}_\ell,\mathbf p_\ell)
        +0.1D(\overline{\mathbf y}_\ell,\mathbf y_\ell)$
        \State Update
        $\{P_{\ell,k},B_{\ell,k},C_{\ell,k},
        \boldsymbol{\theta}_{\ell,k}\}_{
        k\in\{\mathrm g,\mathrm u,\mathrm d\}}$
        using $\mathcal L_{\mathrm{local}}$
    \EndFor
    \State Let $\Theta_\ell^*$ be the checkpoint with the
    lowest validation $\mathcal L_{\mathrm{local}}$
\EndFor
\State \Return $\{\Theta_\ell^*:\ell\in\mathcal I\}$
\end{algorithmic}
\end{algorithm*}

\subsection{Stage II: End-to-End Knowledge Distillation}
\label{sec:end_to_end_distillation}

In Stage~I, each compressed MLP is reconstructed independently using layer inputs from the uncompressed model. After these modules are inserted into the student model, changes introduced by earlier compressed layers may alter the inputs received
by later MLPs. Reconstruction errors may propagate and accumulate across layers. Therefore, Stage~II is introduced to recover the performance of the complete compressed model through end-to-end knowledge distillation. The Stage~II distillation procedure is summarized in Algorithm~\ref{alg:stage2}.

In Stage~II, each selected MLP in the original model is first replaced with its corresponding compressed module obtained in Stage~I to form the student model $S$.
All parameters of the teacher model $T$ are then frozen, and the same token sequences are fed to both models. The student model is optimized end-to-end under supervision from the teacher's output probability distributions and the ground-truth token labels
by minimizing the final end-to-end objective
\begin{equation}
\mathcal{L}_{\mathrm{E2E}}
=\lambda\mathcal{L}_{\mathrm{CE}}
+(1-\lambda)\mathcal{L}_{\mathrm{KD}},
\label{eq:e2e_objective}
\end{equation}
where $\mathcal{L}_{\mathrm{CE}}$ is the next-token
cross entropy loss, $\mathcal{L}_{\mathrm{KD}}$ is
the Top-$K$ knowledge distillation loss, and
$\lambda\in[0,1]$ controls their relative weights. Notably, only the parameters of the compressed MLPs are updated during end-to-end optimization, and the remaining student parameters are kept frozen.

Let $(x_1,\ldots,x_L)$ denote a training sequence, where $x_{i+1}$ is the ground-truth next token following the prefix $x_{\leq i}$.
For Stage~II, let $\Omega$ be the set of positions $i$ for which the next token target $x_{i+1}$ is valid and unmasked.
\begin{equation}
\mathcal{L}_{\mathrm{CE}}=-\frac{1}{|\Omega|}\sum_{i\in\Omega}\log p_S(x_{i+1}\mid x_{\leq i}).
\label{eq:ce_loss}
\end{equation}

To restrict the distillation objective to a small set of candidate tokens while retaining the ground-truth target, we construct a candidate set for each position
\begin{equation}
\mathcal{C}_i=\operatorname{TopK}\left(p_T(\cdot\mid x_{\leq i}),K\right)\cup\{x_{i+1}\}.
\label{eq:topk_candidate_set}
\end{equation}
The teacher and student logits are temperature-scaled and renormalized over $\mathcal{C}_i$. For $M\in\{T,S\}$, we define
\begin{equation}
p_{M,i}^{(\tau)}(v)=\frac{\exp\left(z_{M,i,v}/\tau\right)}{\sum_{u\in\mathcal{C}_{i}}\exp\left(z_{M,i,u}/\tau\right)},\qquad v\in\mathcal{C}_{i}.
\label{eq:restricted_distribution}
\end{equation}
The Top-$K$ distillation loss is
\begin{equation}
\mathcal{L}_{\mathrm{KD}}=\frac{\tau^{2}}{|\Omega|}\sum_{i\in\Omega}\sum_{v\in\mathcal{C}_{i}}p_{T,i}^{(\tau)}(v)\log\frac{p_{T,i}^{(\tau)}(v)}{p_{S,i}^{(\tau)}(v)}.
\label{eq:topk_kd_loss}
\end{equation}

\begin{algorithm*}[h]
\caption{Stage II: End-to-End Distillation and Gain Folding}
\label{alg:stage2}
\begin{algorithmic}[1]
\Require Frozen teacher $T$; student $S$ initialized from
the original model; target layers $\mathcal I$;
Stage-I checkpoints $\{\Theta_\ell^*\}_{\ell\in\mathcal I}$;
distillation data $\mathcal D_{\mathrm{KD}}$ with a
held-out validation split; training steps $N_2$;
$K=64$, $\tau=2$, and $\lambda=0.5$
\Ensure Compressed student $S_{\mathrm{deploy}}$
without photonic inference overhead

\State Replace the target MLPs of $S$ with
$\{\Theta_\ell^*\}_{\ell\in\mathcal I}$
and freeze all other student parameters
\For{$t=1$ to $N_2$}
    \State Run $T$ and $S$ on the same distillation batch
    \State Let $\Omega$ be the positions with valid, unmasked next-token targets
    \State Compute $\mathcal L_{\mathrm{CE}}$
    over $\Omega$ using Eq.~\eqref{eq:ce_loss}
    \For{each $i\in\Omega$}
        \State $\mathcal C_i\gets \operatorname{TopK} \bigl(p_T(\cdot\mid x_{\leq i}),K\bigr) \cup\{x_{i+1}\}$
    \EndFor
    \State Compute the temperature-scaled distributions
    over $\mathcal C_i$ using
    Eq.~\eqref{eq:restricted_distribution}
    \State Compute $\mathcal L_{\mathrm{KD}}$
    using Eq.~\eqref{eq:topk_kd_loss}
    \State $\mathcal L_{\mathrm{E2E}}
    \gets \lambda\mathcal L_{\mathrm{CE}}
    +(1-\lambda)\mathcal L_{\mathrm{KD}}$
    \State Update the trainable parameters of the
    compressed MLPs using $\mathcal L_{\mathrm{E2E}}$
\EndFor
\State Restore the student checkpoint with the lowest
validation $\mathcal L_{\mathrm{E2E}}$
\For{each $\ell\in\mathcal I$ and
$k\in\{\mathrm g,\mathrm u,\mathrm d\}$}
    \State Compute the final input-independent
    gain $\mathbf g_{\ell,k}$
    \State $B_{\ell,k}^{\mathrm{fused}}
    \gets\operatorname{diag}(\mathbf g_{\ell,k})
    B_{\ell,k}$
    \State Implement the projection using the low-rank
    pair $(P_{\ell,k},B_{\ell,k}^{\mathrm{fused}})$
\EndFor
\State Remove the photonic circuits and mapping matrices
\State Let $S_{\mathrm{deploy}}$ be the resulting student
\State \Return $S_{\mathrm{deploy}}$
\end{algorithmic}
\end{algorithm*}

\subsection{Gain Folding for Classical Inference}
\label{sec:gain_folding}

During training, photonic gating scales the low-rank
bottleneck activations using the gain vectors defined
in Eq.~\eqref{eq:channel_gain}.
After training, the circuit parameters and mapping
matrices are fixed.
Because the circuits receive fixed input states rather
than token-dependent inputs, the probability vector
$\mathbf z_k$ and the corresponding gain vector
$\mathbf g_k$ are constant for each projection $k$.

For an input $\mathbf x$ to projection $k$, the forward
computation can be rewritten as
\begin{equation}
P_k\left(\mathbf g_k\odot B_k\mathbf x\right)
=
P_k\operatorname{diag}(\mathbf g_k)B_k\mathbf x,
\label{eq:gain_matrix_form}
\end{equation}
where $\operatorname{diag}(\mathbf g_k)$ is the diagonal
matrix whose diagonal entries are the components of
$\mathbf g_k$.
We absorb the gains into $B_k$ by defining the fused
low-rank factor
\begin{equation}
B_k^{\mathrm{fused}}
=
\operatorname{diag}(\mathbf g_k)B_k.
\label{eq:fused_matrix}
\end{equation}
The projection can then be evaluated as
\begin{equation}
P_k\left(\mathbf g_k\odot B_k\mathbf x\right)
=
P_kB_k^{\mathrm{fused}}\mathbf x.
\label{eq:fused_projection}
\end{equation}

This folding is applied independently to the gate,
up, and down projections of each compressed MLP.
It is an exact algebraic transformation that preserves
the function of the trained compressed MLP and requires
no additional fine-tuning.

After folding, each compressed projection retains only
$P_k$ and $B_k^{\mathrm{fused}}$.
The mapping matrix $C_k$, the parameters of photonic quantum circuit,
and the photonic quantum hardware are no longer required
for inference.
The compressed projections are therefore evaluated
using standard classical low-rank matrix multiplications.

This folding relies on the input-independent
gain formulation and does not work when
the gain vectors are conditioned on the current input.

\subsection{Parameter Count and Computational Complexity}
\label{sec:complexity}

Ignoring bias terms, an original SwiGLU MLP with projection input dimension $d$ and output dimension $m$ contains three dense projection matrices, giving
\begin{equation}
N_{\mathrm{dense}}=3dm
\label{eq:dense_parameter_count}
\end{equation}
parameters. Our method uses rank $r$ for the gate and up projections and rank $2r$ for the down projection.
After gain folding, the number of parameters in one compressed MLP is
\begin{equation}
\begin{aligned}
N_{\mathrm{deploy}}=2r(d+m)+2r(d+m)=4r(d+m).
\end{aligned}
\label{eq:deployed_parameter_count}
\end{equation}

Let $q$ denote the number of photonic output modes. The three mapping matrices $C_k$ contain $4rq$ parameters in total.
If $N_{\mathrm{ph}}$ denotes the total number of trainable parameters in the three photonic circuits, the training-time parameter count of one compressed MLP is
\begin{equation}
N_{\mathrm{train}}=4r(d+m)+4rq+N_{\mathrm{ph}}.
\label{eq:training_parameter_count}
\end{equation}
The mapping matrices and photonic parameters are used only during training and are excluded from the deployed model. The parameter reduction ratio for one deployed MLP is therefore
\begin{equation}
R_{\mathrm{param}}=1-\frac{4r(d+m)}{3dm}.
\label{eq:parameter_reduction_ratio}
\end{equation}

If the MLPs in the selected layers $\mathcal I$ share input dimension $d$, output dimension $m$, and base rank $r$, the total number of student parameters
after replacing these MLPs is
\begin{equation}
N_{\mathrm{student}}=N_{\mathrm{4B}}-|\mathcal I|\left[3dm-4r(d+m)\right],
\label{eq:student_parameter_count}
\end{equation}
where $N_{\mathrm{4B}}$ is the parameter count of the uncompressed 4B model.

In terms of computation, the three dense projections require approximately $3dm$ multiply--accumulate operations per token. The gain-folded low-rank projections require
\begin{equation}
\operatorname{MACs}_{\mathrm{deploy}}=4r(d+m)
\label{eq:deployed_macs}
\end{equation}
multiply--accumulate operations per token, excluding lower-order element-wise operations in SwiGLU.
The theoretical reduction in the dominant linear computation therefore matches the parameter reduction in Eq.~\eqref{eq:parameter_reduction_ratio}.

During training, each target layer additionally evaluates its photonic circuits and the mappings $C_k\mathbf{z}_k$.
Because the photonic gains are input-independent, each module can compute one gain vector per forward pass and share it across all tokens, without evaluating the circuit separately for each token.
After gain folding, the additional photonic computation is completely removed from inference.
\section{Experimental Setup}
\label{sec:experiments}

This section describes the experimental setup used to evaluate QuLoC. We first explain how photonic feature vectors are constructed
using 16-mode and 4-mode photonic quantum circuits in
Section~\ref{sec:experimental_generation_zk}.
With these feature constructions established, we specify the models, teacher configurations, training data, and optimization settings
in Section~\ref{sec:training_settings}.
We then introduce the baselines and controlled comparisons in Section~\ref{sec:baselines} to assess the effects of low-rank training, local reconstruction, and gating based on measured
photonic responses. Finally, Section~\ref{sec:evaluation} defines the compression
and performance metrics and describes the downstream tasks
and evaluation protocols used to compare the resulting models.

\subsection{Constructing the Photonic Feature Vector $\mathbf z_k$}
\label{sec:experimental_generation_zk}

We use two photonic quantum implementations to construct the vector
$\mathbf z_k\in\mathbb R^{16}$ in the gain mapping
$\mathbf g_k=\mathbf 1+\alpha\tanh(C_k\mathbf z_k)$.
The first evaluates the output probabilities of a quantum circuit with
16 spatial modes numerically, using a state containing one photon.
The second combines experimental detection probabilities from an
ensemble of quantum circuits, each with four spatial modes, through a
trainable classical transformation.
We distinguish trainable optical phases from trainable classical
parameters.
The optical phase parameters for projection $k$ are collected in
$\boldsymbol\theta_k$.
The matrix $C_k$ is a trainable classical map.
The experimental implementation also includes classical expansion
coefficients and an affine transformation, defined below.
In the following, we describe one projection and omit the index $k$.
The same construction is applied separately to the gate, up, and
down projections.
Hats distinguish operators on quantum states from classical matrices
in both implementations.

\paragraph{Numerical implementation with 16 spatial modes.}
Let $\hat U(\boldsymbol\theta)$ denote the evolution operator of
a parameterized optical interferometer, and let $|\psi_{\mathrm{in}}\rangle$
be the fixed initial state containing one photon.
In this implementation, $\boldsymbol\theta$ consists of the
trainable circuit phases.
The basis state $|j\rangle$ denotes a photon in mode $j$, with all
other modes unoccupied.
On the subspace containing one photon, $\hat U$ has a
$16\times16$ matrix representation.
The components of $\mathbf z$ are
\begin{equation}
z_j=\left|\langle j|\hat U(\boldsymbol\theta)
|\psi_{\mathrm{in}}\rangle\right|^2,
\qquad j=1,\ldots,16.
\label{eq:setup_sixteen_mode_z}
\end{equation}
These probabilities describe mutually exclusive detection outcomes
and sum to one. The circuit architecture and the corresponding unitary transformation are described in Appendix~\ref{sec:photonic_preliminaries}.

\paragraph{Experimental implementation with 4 spatial modes.}
The experimental implementation uses a \emph{quantum chip ensemble}
with seven distinct circuit configurations, each acting on four
spatial modes.
Each spatial mode corresponds to a distinct optical propagation
channel.
The joint state is described in the Fock basis
$|n_1,n_2,n_3,n_4\rangle$, where $n_j$ is the number of photons in
channel $j$.

We prepare light in a squeezed vacuum state, ($ |\psi\rangle=S(\xi)|0\rangle $), which is a coherent superposition of even-photon-number Fock states with a photon-number distribution governed by the squeezing parameter. The state is then injected into an interferometer with a programmable phase $\phi$. Here, $\phi$ is a scalar phase angle, measured in radians and
defined modulo $2\pi$.
To describe one member of the ensemble, we omit its configuration
index and write the ideal transformation as
\begin{equation}
\hat U(\phi)=\hat U_2\hat E(\phi)\hat U_1.
\label{eq:setup_four_mode_circuit}
\end{equation}
Here, $\hat U$ acts on the Fock space of the four optical modes.
The operators $\hat U_1$ and $\hat U_2$ are fixed transformations that mix
the optical modes, and $\hat E(\phi)$ applies the phase $\phi$.
Each configuration specifies its own fixed transformations
$\hat U_1$ and $\hat U_2$ and the layout of the phase operation.
Thus, choosing a configuration selects a circuit, whereas changing
$\phi$ varies the phase within that circuit.
Varying this phase changes the interference and therefore the
probabilities of the detection outcomes.
For the prepared optical state $\hat\rho$, these probabilities are
given by the Born rule:
\begin{equation}
\pi_s(\phi)=\operatorname{Tr}\!\left[
\hat\Pi_s\hat U(\phi)\hat\rho\hat U^\dagger(\phi)\right],
\label{eq:setup_four_mode_probability}
\end{equation}
where $\hat\Pi_s$ is the POVM element associated with outcome $s$.


\paragraph{Detection outcomes.}
Output light is measured using superconducting nanowire detectors.
A \emph{click} is an electrical signal indicating that a detector
has registered an event within a detection window.
In ideal detection without noise, it indicates at least one
detected photon.
A threshold detector produces the same binary signal when one or
several photons are detected within that window, so one click does
not determine the exact photon number.
To obtain partial information about photon number, the light from
each spatial mode is distributed between two detection channels.
The two channels yield zero, one, or two clicks.
Two clicks indicate that both detectors have responded; several
photons can still produce only one or two clicks.
This provides approximate resolution of photon number.
A joint outcome is recorded as
$s=(s_1,s_2,s_3,s_4)$, with $s_j\in\{0,1,2\}$.
Here, $s_j$ counts the clicks in the two detection channels
associated with spatial mode $j$.
For example, $(1,1,0,0)$ denotes one click from each of the first
two spatial modes and none from the remaining modes.
With this recording convention, there are $3^4=81$ possible patterns,
including the pattern with no clicks.
Clicks registered in different detection channels within a common
time window are grouped into coincidence events.
We observed coincidence records with up to six clicks across the
detection channels.

\paragraph{Probability response curves.}
A \emph{response curve} is the probability of one fixed detection
pattern as a function of the phase in one fixed circuit:
$\phi\mapsto\pi_s(\phi)$.
For example, fixing a circuit and selecting $s=(1,1,0,0)$ gives
one curve that describes how the probability of this pattern
changes as $\phi$ is scanned.
At each phase, repeated measurements give the frequency of the
pattern, which estimates its probability.
Throughout the scan, $\hat U_1$, $\hat U_2$, and the remaining
optical settings are fixed.
Another detection pattern, or the same pattern in a different
ensemble member, defines another curve.

Across the seven fixed quantum circuit configurations used in this work,
we retain 373 response curves in total.\footnote{We retain patterns
with at least one click and at least 100 recorded events, giving
44--63 patterns per configuration.
Following phase calibration, the curves are aligned to 100 phase
points over $[0,2\pi)$ by interpolation, with Fourier fitting used
to complete uncovered intervals when necessary.}
These curves supply the nonlinear functions that are combined to
construct the model features.
Within each configuration, we normalize the probabilities over
the retained pattern set $\mathcal R$:
\begin{equation}
\widetilde\pi_s(\phi)
=\frac{\pi_s(\phi)}{\sum_{s'\in\mathcal R}\pi_{s'}(\phi)},
\qquad s\in\mathcal R.
\label{eq:setup_conditional_response}
\end{equation}
Thus, these responses are conditional probabilities for the
retained outcomes.
Together, the normalized curves form a dictionary of nonlinear
basis functions; linear independence of all curves is not required.

\paragraph{Construction of the feature vector.}
Let $\widetilde{\boldsymbol\pi}(\phi)\in\mathbb R^{373}$ collect the conditional
probabilities $\widetilde\pi_s(\phi)$ from all seven configurations.
In this implementation, the trainable phase vector is
$\boldsymbol\theta=(\phi_1,\ldots,\phi_{16})^\top$.
The scalar $\phi$ labels the phase variable of a response curve;
its learned value $\phi_j$ is used to construct feature $j$.
Evaluating the 373 curves at $\phi_j$ gives 373 scalar values.
Their weighted sum, together with an offset, produces one feature:
\begin{equation}
f_j=\delta_j+\boldsymbol\gamma_j^\top
\widetilde{\boldsymbol\pi}(\phi_j),
\qquad j=1,\ldots,16,
\label{eq:setup_four_mode_readout}
\end{equation}
where $\delta_j$ is a scalar offset and
$\boldsymbol\gamma_j\in\mathbb R^{373}$ contains the weights assigned to the
measured outcomes.
Repeating this construction with 16 trainable phase values and
corresponding weights produces
$\mathbf f=(f_1,\ldots,f_{16})^\top$.

A trainable affine transformation then gives
\begin{equation}
\mathbf z=A\mathbf f+\mathbf b,
\qquad
A\in\mathbb R^{16\times16},\quad
\mathbf f,\mathbf b\in\mathbb R^{16}.
\label{eq:setup_four_mode_z}
\end{equation}
The resulting vector contains 16 real features and is not subject
to probability normalization. For each feature, seven trainable coefficients
$\boldsymbol{\beta}_j\in\mathbb R^7$ determine the offset
$\delta_j$ and weight vector $\boldsymbol{\gamma}_j$ through a
fixed linear map.
The coefficients $\boldsymbol{\beta}_j$, the affine parameters
$A$ and $\mathbf b$, and the matrix $C$ are classical trainable
parameters; the trainable optical phases are collected in
$\boldsymbol\theta=(\phi_1,\ldots,\phi_{16})^\top$.

The gate, up, and down projections share the same measured response
curves and fixed calibration maps, while each has its own optical
phase vector $\boldsymbol\theta_k$ and classical parameters.
These parameters are jointly optimized through the task loss,
with the optical transformations $\hat U_1$ and $\hat U_2$ and the
calibration maps held fixed.
All learned parameters are shared across tokens, and the phases
do not encode the MLP input $\mathbf x$.
Consequently, $\mathbf z_k$ and $\mathbf g_k$ are independent of
the current token.

\subsection{Models, Data, and Training Settings}
\label{sec:training_settings}

\paragraph{Teacher-Student Configuration} Our training procedure uses different teacher configurations in the two stages.
Stage~I reconstructs each compressed MLP using the corresponding
original MLP as the reference, requiring matching input and
output dimensions. We use a complete and frozen Qwen3.5-4B as the Stage~I teacher, denoted by $\mathcal{M}_{T}^{4\mathrm{B}}$, and initialize the student $\mathcal{M}_{S}$ from its weights.
Given a set of target layers $\mathcal{I}$, 
we replace the SwiGLU MLPs in it with the proposed low-rank SwiGLU MLPs.
The token embeddings, attention modules, normalization layers, output head, and non-target MLPs of the student remain frozen.
Stage~I directly uses the layer inputs and outputs of the 4B teacher to supervise the corresponding compressed student modules.

After layerwise reconstruction, the selected modules are assembled into a single compressed student model.
For Stage~II, the frozen teacher can be either Qwen3.5-4B or Qwen3.5-27B:
\begin{equation}
\mathcal{M}_{T}^{\mathrm{E2E}}\in\left\{\mathcal{M}_{T}^{4\mathrm{B}},\mathcal{M}_{T}^{27\mathrm{B}}\right\}.
\label{eq:stage2_teacher}
\end{equation}
Using the 4B teacher corresponds to same-scale distillation, whereas the 27B teacher transfers knowledge from a larger model to the compressed 4B student. Stage~II matches next-token distributions in the shared vocabulary
rather than intermediate representations. This allows either
teacher to supervise the student without requiring matching
hidden dimensions. The teacher remains frozen in both settings, and only the compressed parameters in the target student layers are updated.

For LLaMA-7B, the frozen uncompressed LLaMA-7B model serves as the teacher in both stages, providing a consistent reference for local reconstruction and end-to-end distillation. This setting allows us to evaluate how well QuLoC preserves
and recovers the original model's performance without
introducing supervision from a larger teacher.

\begin{figure}
    \centering
    \includegraphics[width=\linewidth]{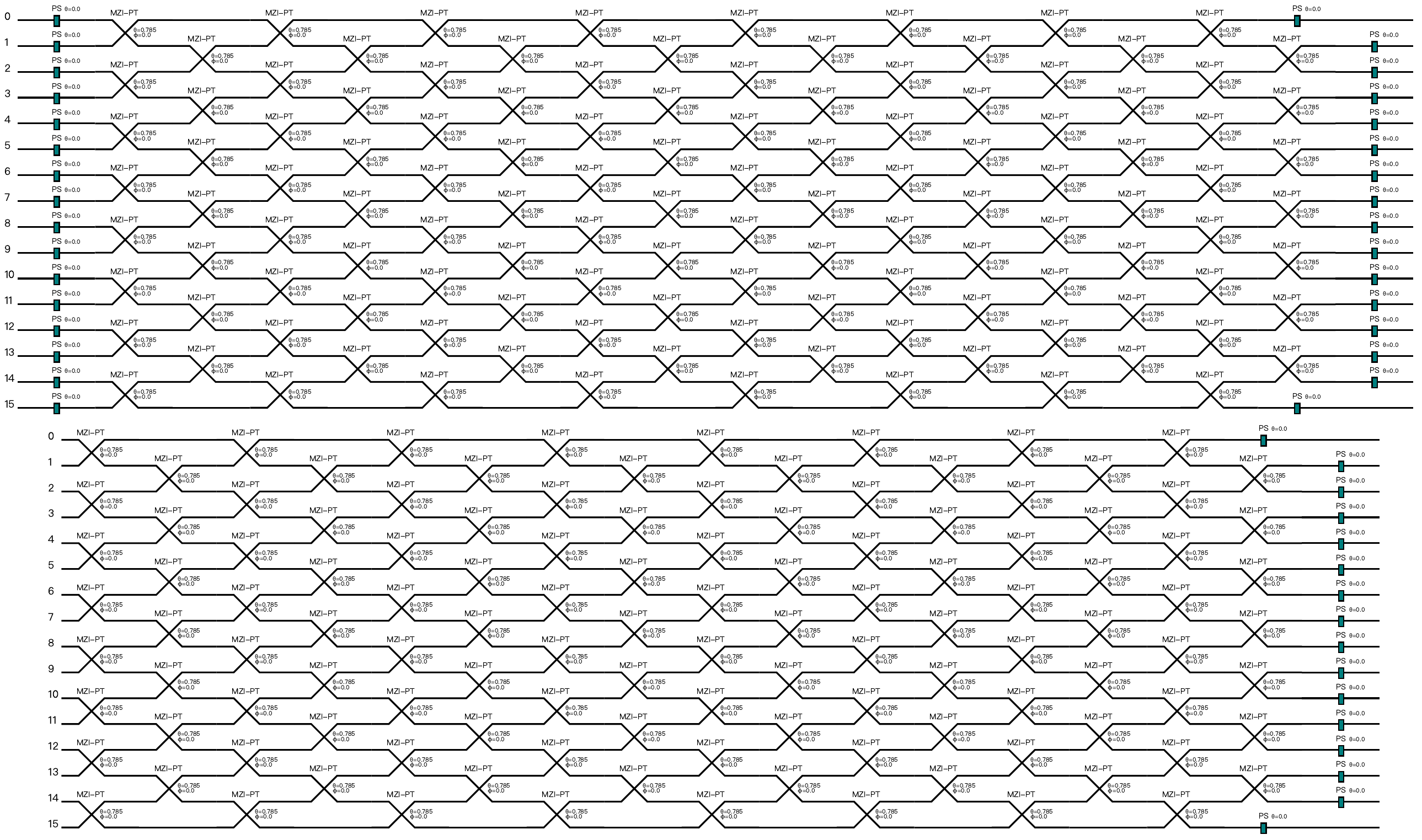}
    \caption{Architecture of the 16 mode photonic quantum circuit used in each compressed projection module. An input independent phase encoding of a single photon state is processed by two cascaded trainable Clements interferometer meshes, separated by a trainable inter-mesh phase layer. For visual clarity, the two cascaded meshes are drawn in separate rows: the output of the first row passes through the inter-mesh phase layer before entering the second row. The parameter values shown for trainable components are their initial values and are optimized during training. The photon-number probabilities at the output modes form the feature vector used to generate the low rank channel gates. The gate, up, and down projection modules employ independent copies of this circuit.}
    \label{fig:double_clements}
\end{figure}

The main Qwen3.5-4B experiment replaces the MLPs in layers 8--23 with rank-$384$ gate and up projections and rank-$768$ down projections.
The Stage~I ablation using a 16 modes photonic quantum circuit instead compresses the last four MLPs (layers 28--31, indexed from zero), using rank $512$ for the gate and up projections and rank $1024$ for the down projection.
Each compressed projection uses an independent photonic circuit containing two trainable Clements interferometer meshes, as illustrated in Figure~\ref{fig:double_clements}.
The low-rank factors $P$ and $B$ are initialized by truncated SVD, while $C$ and the parameters of the photonic circuit are initialized from a zero-mean Gaussian distribution with a standard deviation of $0.1$.
Both training stages use UltraChat, with non-overlapping training and validation subsets.

Stage~I is optimized for 1,500 steps with a batch size of 64, followed by 1,000 steps of Stage~II training with a batch size of 24, using Adam on an NVIDIA RTX PRO 6000 BLACKWELL.
The learning rates for the different parameter groups are
\begin{equation}
\eta_C=2\times10^{-4}, \qquad \eta_Q=1\times10^{-3}, \qquad \eta_{PB}=5\times10^{-5}.
\label{eq:stage1_learning_rates}
\end{equation}
The local reconstruction loss is evaluated on the validation set every 200 steps, and the checkpoint with the lowest validation loss is retained.

In Stage~II, the compressed parameters in all target layers are jointly optimized, while the remaining student parameters are frozen.
The distillation candidate set contains the teacher's top $K=64$ tokens and the ground-truth token.
We set the distillation temperature to $\tau=2$ and the loss-mixing coefficient to $\lambda=0.5$.

\subsection{Baselines and Controlled Comparisons}
\label{sec:baselines}

We compare QuLoC with the following baselines:
\begin{itemize}
    \item \textbf{Dense}: the uncompressed Qwen3.5-4B model.
    \item \textbf{Truncated SVD}: the selected projections are replaced by their truncated-SVD approximations without training.
    \item  \textbf{Optimized SVD}: the target projections are initialized by truncated SVD, and their low-rank factors are optimized through both training stages, with channel gains fixed to one.
    \item \textbf{Swift-SVD}: a state-of-the-art classical low-rank compression method used as a baseline.
\end{itemize}
Optimized SVD and QuLoC use the same target layers, ranks, data splits, and two-stage optimization budget.

We perform three controlled comparisons:
\begin{itemize}
    \item \textbf{Low-rank training}: Truncated SVD and Optimized SVD separate direct truncation from subsequent two-stage optimization.
    \item \textbf{Stage~I reconstruction}: direct Stage~II training is compared with Stage~I followed by Stage~II, starting from the same initialization and using the same Stage~II budget.
    \item \textbf{Photonic gating}: we compare Optimized SVD with a variant whose gating features come from photonic quantum circuits.
\end{itemize}

The 27B model is used only as a Stage-II teacher and a performance reference.
Its parameter count and inference efficiency are not treated as directly comparable to those of the compressed 4B student.

\subsection{Evaluation Metrics and Downstream Tasks}
\label{sec:evaluation}

For Qwen3.5-4B, we report the reduction in parameters of language model relative to that of the uncompressed Qwen3.5-4B model, excluding the vision encoder.
All compressed models in the main comparison target approximately the same parameter reduction of $20.05\%$.
For LLaMA-7B, we report parameter retention ratios of $\rho=0.8$ and $\rho=0.6$, along with the memory required to store the model weights at each ratio.

We report perplexity (PPL) on WikiText-2 and C4 for both model families.
Let $N_{\mathrm{tok}}$ denote the total number of valid next-token prediction targets. For target position $i$, let $w_i$
be the target token and $c_i$ its available context within the evaluation sequence.
PPL is computed from the mean next-token negative log-likelihood:
\begin{equation}
\operatorname{PPL}=\exp\left(-\frac{1}{N_{\mathrm{tok}}}\sum_{i=1}^{N_{\mathrm{tok}}}\log p_{\theta}(w_i\mid c_i)\right).
\label{eq:evaluation_ppl}
\end{equation}

For our evaluations, documents are concatenated and tokenized before being partitioned into non-overlapping sequences of 2,048 tokens.
The C4 evaluation uses 2,000 validation documents selected after shuffling with seed 7.
All models within a model family are evaluated using exactly the same token stream and tokenizer.
Because the two model families use different tokenizers, PPL values are compared within, rather than across, model families.
The LLaMA-7B entries marked with $\dagger$ are values reported by Swift-SVD~\cite{swift-svd}.

We evaluate zero-shot downstream performance using
\texttt{lm-evaluation-harness} on ARC-Easy, PIQA,
OpenBookQA, WinoGrande, HellaSwag, and MathQA.
For ARC-Easy, PIQA, OpenBookQA, HellaSwag, and MathQA,
we report the harness-defined length-normalized accuracy
(\texttt{acc\_norm}).
Let $x$ be an input prompt and $a_j$ its $j$-th candidate answer.
For the student model $S$, the length-normalized candidate score is
\begin{equation}
s_{\mathrm{norm}}(a_j\mid x)=\frac{1}{L(a_j)}\log p_S(a_j\mid x)=\frac{1}{L(a_j)}\sum_{t=1}^{T_j}\log p_S(a_{j,t}\mid x,a_{j,<t}),
\label{eq:normalized_candidate_score}
\end{equation}
where $T_j$ is the number of tokens in $a_j$ and $L(a_j)$ is the length of the candidate string used by the harness.
A prediction is correct when the candidate with the highest score is the ground-truth answer.
For WinoGrande, we report standard accuracy (\texttt{acc}), which selects the candidate with the highest unnormalized log-likelihood.
The macro average is the unweighted arithmetic mean of the six task accuracies.

For the Stage~I ablation, we additionally report the local reconstruction loss.
For Stage~II, we report the cross-entropy (CE), knowledge-distillation (KD), and combined training losses, together with downstream accuracy on PIQA, ARC-Easy, and OpenBookQA.
Photonic quantum hardware experiments are evaluated using the same WikiText-2, C4, and six-task protocols.

\section{Results and Analysis}
\label{sec:analysis}

This section evaluates the proposed method from four perspectives.
We first compare it with classical compression baselines on Qwen3.5-4B.
We then examine whether Stage~I local reconstruction provides a better initialization for subsequent end-to-end optimization.
Next, we evaluate the method on LLaMA-7B at two parameter-retention ratios.
Finally, we examine whether hardware-measured responses can provide features for photonic gating.  

The main comparison uses rank $r=384$ and replaces the MLPs in layers 8--23 of Qwen3.5-4B.
Parameter reduction is computed over the language-model component, excluding the vision encoder.
This configuration reduces the language-model parameter count from 4,205,751,296 to 3,362,696,192, corresponding to a reduction of 20.05\%.
The compression configuration is selected based on the rank and layer-number ablations reported in Appendix~\ref{app:compression_configuration_ablation}.
Specifically, Appendix~\ref{app:rank_ablation} examines the effect of rank under a fixed compression scope, while Appendix~\ref{app:depth_ablation} evaluates the effect of the number of compressed middle MLP layers. Together, these analyses guide the choice of rank and compression
scope by characterizing the trade-off between parameter reduction and model performance.
In addition, Appendix~\ref{app:gating_depth} examines how the gains from photonic gating vary with the number of compressed layers.

\subsection{Comparison with Classical Compression Baselines}
\label{sec:comparison_classical}

The uncompressed Qwen3.5-4B model serves as the performance reference.
We compare three configurations with the same compressed layers, ranks, and deployed parameter count: Truncated SVD, Optimized SVD, and QuLoC.
Truncated SVD replaces the target projections with their truncated-SVD approximations without subsequent training.
Optimized SVD starts from the same factors and trains them through Stage~I and Stage~II while fixing every channel gain to one.
QuLoC follows the same two-stage procedure, using photonic gating features to modulate the low-rank channels.
For both trained configurations, the frozen Qwen3.5-4B model serves as the Stage~I teacher and the frozen Qwen3.5-27B model serves as the Stage~II teacher.

We also include Swift-SVD as a classical compression reference~\cite{swift-svd}.
All compressed models in Table~\ref{tab:main_comparison_qwen} have the same language-model parameter reduction of approximately 20.05\%.
We report token-level perplexity on WikiText-2 and C4 and zero-shot accuracy on ARC-Easy, PIQA, OpenBookQA, WinoGrande, HellaSwag, and MathQA.
Accuracy is length-normalized for all tasks except WinoGrande, for which standard accuracy is used; the macro average covers all six tasks.

\begin{table*}[t]
    \centering
    \setlength{\tabcolsep}{4pt}
    \renewcommand{\arraystretch}{1.35}
    \caption{Performance comparison on Qwen3.5-4B at a matched language-model parameter reduction of approximately 20.05\%. Parameter reduction is computed over the language-model component. WikiText-2 and C4 perplexities are computed at the token level using non-overlapping sequences of 2,048 tokens; C4 uses 2,000 validation documents. ARC-Easy, PIQA, OpenBookQA, HellaSwag, and MathQA use normalized accuracy, whereas WinoGrande uses standard accuracy. Bold values indicate the best result among the compressed models.}
    \label{tab:main_comparison_qwen}
\resizebox{\textwidth}{!}{%
\begin{tabular}{l|cc|cccccc|c}
    \toprule
    \textbf{Model/Method}
    & \multicolumn{2}{c|}{\textbf{PPL} $\downarrow$}
    & \multicolumn{6}{c|}{\textbf{Accuracy} $\uparrow$}
    & \textbf{Average} $\uparrow$ \\
    \cmidrule(lr){2-3}
    \cmidrule(lr){4-9}
    & \textbf{WikiText-2}
    & \textbf{C4}
    & \textbf{ARC-E}
    & \textbf{PIQA}
    & \textbf{OBQA}
    & \textbf{WinoG.}
    & \textbf{HellaS.}
    & \textbf{MathQA}
    & \\
    \midrule

    Qwen3.5-4B
    & 9.570
    & 14.332
    & 0.7551
    & 0.7851
    & 0.4060
    & 0.6969
    & 0.7299
    & 0.4265
    & 0.6332 \\
    \midrule

    Truncated SVD
    & 71.079
    & 80.272
    & 0.5257
    & 0.6540
    & 0.3460
    & 0.5359
    & 0.4361
    & 0.2101
    & 0.4513 \\

    Optimized SVD
    & 14.133
    & 26.027
    & 0.6170
    & 0.6915
    & 0.3540
    & 0.5512
    & 0.4538
    & 0.2449
    & 0.4854 \\

    Swift-SVD
    & 27.646
    & 49.659
    & 0.6103
    & 0.6888
    & 0.3160
    & 0.5643
    & 0.4526
    & \textbf{0.3223}
    & 0.4924 \\

    \textbf{QuLoC (Ours)}
    & \textbf{12.173}
    & \textbf{24.644}
    & \textbf{0.7008}
    & \textbf{0.7209}
    & \textbf{0.3760}
    & \textbf{0.5888}
    & \textbf{0.5705}
    & 0.2848
    & \textbf{0.5403} \\
    \bottomrule
\end{tabular}%
}
\end{table*}

Table~\ref{tab:main_comparison_qwen} shows that QuLoC achieves the lowest perplexity on both WikiText-2 and C4 and the highest macro-average downstream accuracy among the compressed models.
At the matched 20.05\% parameter reduction, QuLoC improves the average accuracy over Swift-SVD from 49.24\% to 54.03\%, a gain of 4.79 percentage points (9.73\% relative).
WikiText-2 perplexity decreases from 27.646 to 12.173, and C4 perplexity decreases from 49.659 to 24.644.
QuLoC performs better on five of the six downstream tasks, while Swift-SVD obtains the higher MathQA accuracy.

The comparison between Truncated SVD and Optimized SVD evaluates the benefit of training the SVD-initialized low-rank factors.
Two-stage optimization reduces WikiText-2 perplexity from 71.079 to 14.133 and C4 perplexity from 80.272 to 26.027.
It also improves the macro-average accuracy from 45.13\% to 48.54\%, a gain of 3.41 percentage points, with improvements on all six downstream tasks.
The specific contribution of Stage~I is examined separately in Section~\ref{sec:results_stage1_effectiveness}.

We assess the effectiveness of photonic gating by comparing QuLoC with Optimized SVD under the same compressed architecture and two-stage training budget.
QuLoC reduces WikiText-2 perplexity from 14.133 to 12.173 and C4 perplexity from 26.027 to 24.644.
The macro-average accuracy rises from 48.54\% to 54.03\%, with improvements on all six tasks.

Overall, optimization after SVD truncation recovers a substantial part of the lost performance, and photonic gating provides a further improvement at the same deployed parameter count.
The compressed model nevertheless remains below the uncompressed Qwen3.5-4B reference in average downstream accuracy.

\subsection{Effectiveness of Stage~I Local Reconstruction}
\label{sec:results_stage1_effectiveness}

We evaluate whether Stage~I local reconstruction provides a better initialization for subsequent end-to-end optimization.
In this experiment, the MLPs in the last four Transformer blocks of Qwen3.5-4B are replaced with photonic-gated low-rank modules.

The gate and up projections use rank $r=512$, while the down projection uses rank $2r$.
The frozen Qwen3.5-4B model serves as the Stage~I teacher, while the frozen Qwen3.5-27B model serves as the Stage~II teacher.
Stage~I is performed for 1,500 optimization steps, followed by 1,000 steps of Stage~II training.

We compare Stage~I $\rightarrow$ Stage~II training with direct Stage~II training.
Both runs start from exactly the same initial parameter values.
The low-rank factors $P$ and $B$ are initialized from the corresponding truncated SVD, while the mapping matrices $C$ and photonic-circuit parameters are sampled from a zero-mean Gaussian distribution with a standard deviation of $0.1$.
The resulting initial parameter state is copied to both runs.

In direct Stage~II training, all compressed parameters are optimized immediately using the end-to-end objective.
In Stage~I $\rightarrow$ Stage~II training, the same initial parameters first undergo layer-wise local reconstruction and are then passed to Stage~II without reinitialization.
The two Stage~II runs use the same random seed, objective, optimization settings, and number of optimization steps.
Their difference is therefore whether the compressed modules undergo local reconstruction before Stage~II.

\begin{figure}[t]
    \centering
    \includegraphics[width=0.8\linewidth]{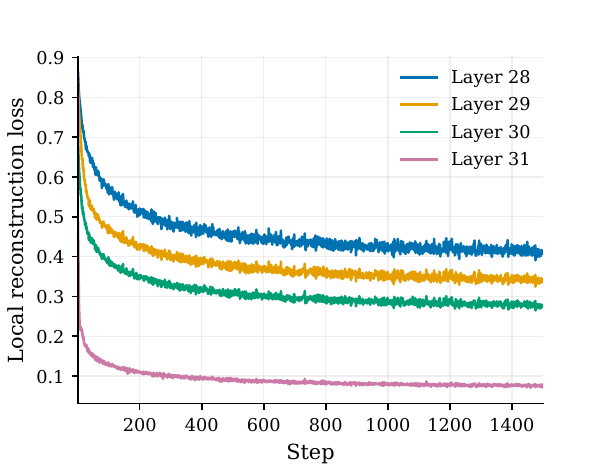}
    \caption{Stage~I local reconstruction loss during the 1,500-step optimization for the four compressed MLPs in the last four Transformer blocks of Qwen3.5-4B. Each curve reports $\mathcal{L}_{\mathrm{local}}$ for one target layer. Lower values indicate better local reconstruction.}
    \label{fig:local_reconstruction_stage1_effect}
\end{figure}

Figure~\ref{fig:local_reconstruction_stage1_effect} shows that the local reconstruction loss decreases consistently for all four target layers during Stage~I optimization.
Most of the reduction occurs during the early optimization steps, followed by more gradual improvement.
The layers reach different loss levels, indicating that their sensitivity to low-rank replacement is layer dependent.
Nevertheless, the consistent reduction across all four layers shows that Stage~I effectively adapts each compressed MLP toward the input--output mapping of its uncompressed counterpart.

\begin{figure}[t]
    \centering
    \includegraphics[width=\linewidth]{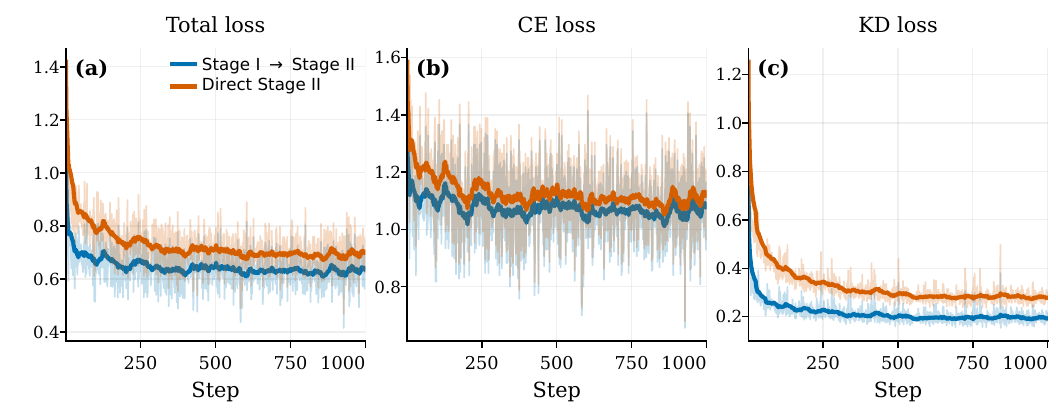}
    \caption{Stage~II optimization with and without Stage~I local reconstruction. Panels (a)--(c) show the total, CE, and KD losses, respectively, where $\mathcal{L}_{\mathrm{E2E}}=0.5\mathcal{L}_{\mathrm{CE}}+0.5\mathcal{L}_{\mathrm{KD}}$. Faint and solid curves denote step-wise losses and 25-step moving averages, respectively.}
    \label{fig:stage2_loss_ablation}
\end{figure}

\begin{table}[t]
    \centering
    \small
    \caption{Stage~II losses with and without Stage~I local reconstruction. Values are averaged over the final optimization steps. Lower values are better.}
    \label{tab:stage1_loss_quantum}
    \begin{tabular}{lcccc}
        \toprule
        \textbf{Loss} & \textbf{Stage~I $\rightarrow$ II} & \textbf{Direct Stage~II} & \textbf{Abs. Red.} & \textbf{Rel. Red.} \\
        \midrule
        Total loss & 0.6392 & 0.7026 & 0.0634 & 9.02\% \\
        KD loss & 0.1927 & 0.2796 & 0.0869 & 31.08\% \\
        CE loss & 1.0856 & 1.1256 & 0.0400 & 3.55\% \\
        \bottomrule
    \end{tabular}
\end{table}

As shown in Figure~\ref{fig:stage2_loss_ablation}, 
Stage~I initialization generally yields lower Stage II losses and reaches a lower final objective. Table~\ref{tab:stage1_loss_quantum} summarizes the losses averaged over the final optimization steps.
Stage~I reduces the total loss from $0.7026$ to $0.6392$.
The KD loss decreases from $0.2796$ to $0.1927$, while the CE loss decreases from $1.1256$ to $1.0856$.

Stage~I first reduces the functional mismatch between each compressed MLP and its uncompressed 4B counterpart.
The resulting initialization is then distilled from the 27B teacher in Stage~II.
The lower KD and CE losses indicate that local reconstruction improves subsequent optimization in this configuration.

\begin{table}[t]
    \centering
    \small
    \caption{
    Zero-shot downstream performance with and without Stage~I local reconstruction. Improvements are reported in percentage points.
    }
    \label{tab:stage1_downstream_quantum}
    \begin{tabular}{lcccc}
        \toprule
        \textbf{Training strategy} & \textbf{PIQA} & \textbf{ARC-Easy} & \textbf{OpenBookQA} & \textbf{Average} \\
        \midrule
        Stage~I $\rightarrow$ Stage~II & 0.7688 & 0.7302 & 0.4020 & 0.6337 \\
        Direct Stage~II & 0.7579 & 0.7222 & 0.3960 & 0.6254 \\
        \midrule
        Improvement & +1.09 pp & +0.80 pp & +0.60 pp & +0.83 pp \\
        \bottomrule
    \end{tabular}
\end{table}

The downstream results in Table~\ref{tab:stage1_downstream_quantum} are consistent with the Stage~II loss analysis.
Compared with direct Stage~II training, Stage~I initialization improves PIQA, ARC-Easy, and OpenBookQA by $1.09$, $0.80$, and $0.60$ percentage points, respectively.
The macro-average accuracy across the three tasks increases from $62.54\%$ to $63.37\%$, yielding a modest but consistent improvement of $0.83$ percentage points.

Under the same Stage II training budget, adding Stage I local reconstruction improves the three-task mean accuracy from 62.54\% to 63.37\%.
Stage I also introduces additional optimization steps, so this experiment establishes the benefit of adding local reconstruction under the present training protocol, rather than a compute-matched advantage over direct end-to-end training.

\subsection{Scalability Evaluation on LLaMA-7B}
\label{sec:llama7b_scalability}

To evaluate the method on a larger model at different compression levels, we apply QuLoC to LLaMA-7B at model parameter-retention ratios of $\rho=0.8$ and $\rho=0.6$.
At $\rho=0.8$, we compress 16 middle MLP layers with rank $r=840$ for the gate and up projections and rank $2r$ for the down projection.
At $\rho=0.6$, we compress 24 middle MLP layers with rank $r=384$ for the gate and up projections and rank $2r$ for the down projection.
The frozen uncompressed LLaMA-7B model serves as the teacher in both training stages.
For comparison, Table~\ref{tab:llama7b_scalability} includes previously reported results, marked with $\dagger$, from Swift-SVD~\cite{swift-svd}.
Swift-SVD uses uniform rank allocation, whereas Swift-SVD* uses dynamic rank allocation based on layer-wise reconstruction loss and layer importance.

\begin{table*}[t]
    \centering
    \setlength{\tabcolsep}{3.2pt}
    \renewcommand{\arraystretch}{1.08}

    \caption{Performance comparison of compression methods on LLaMA-7B at different parameter-retention ratios. The original LLaMA-7B is included as the uncompressed baseline. Rows marked with $\dagger$ are results reported by Swift-SVD \cite{swift-svd}; Swift-SVD and Swift-SVD* use uniform and dynamic rank allocation, respectively. Model-weight memory is shown in parentheses. Lower perplexity (PPL) is better, whereas higher task accuracy and average accuracy are better. Bold values indicate the best result among compressed models at each retention ratio, including ties.}
    \label{tab:llama7b_scalability}

    \resizebox{0.98\textwidth}{!}{%
    \begin{tabular}{c|l|cc|cccccc|c}
        \toprule

        \multirow{2}{*}{Ratio (MEM.)}
        & \multirow{2}{*}{Method}
        & \multicolumn{2}{c|}{PPL ($\downarrow$)}
        & \multicolumn{6}{c|}{Accuracy ($\uparrow$)}
        & \multirow{2}{*}{Avg. ($\uparrow$)} \\

        \cmidrule(lr){3-4}
        \cmidrule(lr){5-10}

        &
        & WikiText-2 & C4
        & ARC-E & PIQA & OBQA & WinoG. & HellaS. & MathQA
        & \\

        \midrule

        1.0 (12.6 GB) & LLaMA-7B$^\dagger$ & 5.68 & 7.34 & 0.76 & 0.79 & 0.34 & 0.70 & 0.57 & 0.27 & 0.57 \\

        \midrule

        \multirow{8}{*}{\begin{tabular}{c}0.8\\(10.1 GB)\end{tabular}}

        & FWSVD$^\dagger$ & 1727 & 1511 & 0.11 & 0.10 & 0.09 & 0.05 & 0.08 & 0.05 & 0.08 \\

        & ASVD$^\dagger$ & 11.14 & 15.93 & 0.53 & 0.68 & 0.29 & 0.64 & 0.41 & 0.17 & 0.45 \\

        & SVD-LLM(W)$^\dagger$ & 7.94 & 15.84 & 0.62 & 0.71 & \textbf{0.31} & 0.61 & 0.45 & 0.21 & 0.49 \\

        & Dobi-SVD(w/o)$^\dagger$ & 8.87 & 10.91 & -- & -- & -- & -- & -- & -- & -- \\

        & Dobi-SVD(w)$^\dagger$ & 8.54 & \textbf{10.01} & 0.63 & 0.72 & 0.30 & 0.62 & 0.46 & 0.20 & 0.49 \\

        & Swift-SVD$^\dagger$ & 7.91 & 11.42 & 0.64 & \textbf{0.73} & 0.26 & \textbf{0.68} & 0.47 & 0.23 & 0.50 \\

        & Swift-SVD*$^\dagger$ & 7.84 & 11.15 & 0.65 & \textbf{0.73} & 0.27 & \textbf{0.68} & 0.48 & 0.23 & 0.51 \\

        & \textbf{QuLoC (Ours)} & \textbf{7.63} & 12.76 & \textbf{0.66} & 0.72 & 0.30 & 0.66 & \textbf{0.52} & \textbf{0.24} & \textbf{0.52} \\

        \midrule

        \multirow{8}{*}{\begin{tabular}{c}0.6\\(7.7 GB)\end{tabular}}
        & FWSVD$^\dagger$ & 18156 & 12847 & 0.05 & 0.05 & 0.06 & 0.02 & 0.00 & 0.03 & 0.04 \\

        & ASVD$^\dagger$ & 1407 & 1109 & 0.11 & 0.13 & 0.08 & 0.09 & 0.08 & 0.08 & 0.10 \\

        & SVD-LLM(W)$^\dagger$ & 13.73 & 75.42 & 0.33 & 0.63 & \textbf{0.25} & 0.55 & 0.40 & 0.12 & 0.38 \\

        & Dobi-SVD(w/o)$^\dagger$ & 14.96 & 24.60 & -- & -- & -- & -- & -- & -- & -- \\

        & Dobi-SVD(w)$^\dagger$ & 13.54 & 23.54 & 0.45 & 0.64 & 0.22 & 0.58 & 0.36 & 0.18 & 0.41 \\

        & Swift-SVD$^\dagger$ & 13.42 & 23.32 & 0.49 & 0.66 & 0.21 & \textbf{0.62} & 0.37 & \textbf{0.22} & 0.43 \\

        & Swift-SVD*$^\dagger$ & \textbf{13.29} & \textbf{21.92} & \textbf{0.51} & \textbf{0.67} & 0.23 & \textbf{0.62} & 0.38 & \textbf{0.22} & \textbf{0.44} \\

        & \textbf{QuLoC (Ours)} & 17.12 & 27.64 & 0.49 & 0.65 & \textbf{0.25} & 0.60 & \textbf{0.42} & 0.21 & \textbf{0.44} \\

        \bottomrule
    \end{tabular}%
    }
\end{table*}

At $\rho=0.8$, QuLoC attains an average downstream accuracy of 52\%, compared with 51\% for Swift-SVD*.
It achieves the highest accuracy among the compressed methods on ARC-Easy, HellaSwag, and MathQA.
Its WikiText-2 perplexity is lower than that of Swift-SVD* (7.63 versus 7.84), whereas its C4 perplexity is higher (12.76 versus 11.15).
Thus, at this retention ratio, the average accuracy gain does not coincide with an improvement on every language-modeling metric.

At $\rho=0.6$, QuLoC and Swift-SVD* both report 44\% average accuracy at the precision shown in Table~\ref{tab:llama7b_scalability}.
QuLoC obtains the highest HellaSwag accuracy and ties for the highest OpenBookQA accuracy among the compressed models. Its perplexity is higher than that of Swift-SVD* on both WikiText-2 (17.12 versus 13.29) and C4 (27.64 versus 21.92).
These results indicate competitive downstream accuracy across the two tested compression settings, but do not show a consistent advantage across all metrics for LLaMA-7B.

\subsection{Experiments with 4-mode photonic quantum hardware}
\label{sec:results_real_device_photonic_gating}

All experiments presented thus far use a numerical implementation of the photonic quantum circuit with 16 spatial modes. In this subsection, we further evaluate the proposed method using experimentally measured data from a 4-mode photonic quantum device. Unlike the preceding simulations, the photonic features used here are constructed from hardware-measured responses, providing an experimental validation of the proposed photonic gating mechanism.

We conduct an auxiliary evaluation using the photonic feature vector constructed from a response pool measured on a photonic quantum device.
The experiment compresses the MLPs in layers 8--23 of Qwen3.5-4B with rank $r=384$ for the gate and up projections and rank $2r$ for the down projection.
The frozen Qwen3.5-4B model serves as the Stage~I teacher, and the frozen Qwen3.5-27B model serves as the Stage~II teacher.
We compare QuLoC using Photonic Quantum Hardware with Optimized SVD under the same compression configuration and evaluate both on WikiText-2, C4, and six zero-shot downstream tasks.

\begin{table*}[t]
    \centering
    \setlength{\tabcolsep}{5pt}
    \renewcommand{\arraystretch}{1.25}
    \caption{Comparison of Optimized SVD and QuLoC using Photonic Quantum Hardware on Qwen3.5-4B. Both methods use rank 384 and compress layers 8--23 of the same Qwen3.5-4B backbone. The hardware model uses photonic features collected from the physical device. WikiText-2 and C4 perplexities are computed at the token level using non-overlapping sequences of 2,048 tokens; C4 uses 2,000 validation documents. ARC-Easy, PIQA, OpenBookQA, HellaSwag, and MathQA use normalized accuracy, whereas WinoGrande uses standard accuracy. Accuracy values are reported as fractions. Bold values indicate the better result.}
    \label{tab:real_device}
    \resizebox{\textwidth}{!}{%
    \begin{tabular}{l|cc|cccccc|c}
        \toprule
        \textbf{Method}
        & \multicolumn{2}{c|}{\textbf{PPL} $\downarrow$}
        & \multicolumn{6}{c|}{\textbf{Accuracy} $\uparrow$}
        & \textbf{Average} $\uparrow$ \\
        \cmidrule(lr){2-3}
        \cmidrule(lr){4-9}
        & \textbf{WikiText-2}
        & \textbf{C4}
        & \textbf{ARC-E}
        & \textbf{PIQA}
        & \textbf{OBQA}
        & \textbf{WinoG.}
        & \textbf{HellaS.}
        & \textbf{MathQA}
        & \\
        \midrule

        Optimized SVD
        & 14.133
        & 26.027
        & 0.6170
        & 0.6915
        & 0.3540
        & 0.5512
        & 0.4538
        & 0.2449
        & 0.4854 \\

        QuLoC (Hardware)
        & \textbf{12.734}
        & \textbf{25.350}
        & \textbf{0.6692}
        & \textbf{0.7133}
        & \textbf{0.3680}
        & \textbf{0.5975}
        & \textbf{0.5630}
        & \textbf{0.2831}
        & \textbf{0.5323} \\
        \bottomrule
    \end{tabular}%
    }
\end{table*}

Table~\ref{tab:real_device} shows that the QuLoC using Photonic Quantum Hardware reduces WikiText-2 perplexity from 14.133 to 12.734 and C4 perplexity from 26.027 to 25.350 relative to Optimized SVD.
Its average downstream accuracy increases from 48.54\% to 53.23\%, a gain of 4.69 percentage points, with improvements on all six tasks.
The largest improvement is on HellaSwag (10.92 percentage points), followed by ARC-Easy (5.22 points) and WinoGrande (4.63 points).

\section{Discussion}
\label{sec:discussion}

This section interprets the experimental results, examines
the roles of the key components of QuLoC, and discusses
the factors that influence compression performance and
the limitations of the method.

\subsection{Interpreting the Contribution of Photonic Gating}
\label{sec:discussion_photonic_gating}

The comparison between QuLoC and the unit-gate low-rank
baseline is intended to assess the contribution of
photonic gating.
Specifically, we consider whether the gain vectors
derived from photon detection probabilities provide
additional flexibility during low-rank MLP reconstruction
and how they modulate the contributions of individual
low-rank components.
Under matched compression and training settings, any
observed performance difference should be interpreted
as the effect of the gating
parameterization, rather than as direct evidence of
a quantum computational advantage.

\subsection{Mechanism of the Two-Stage Training Strategy}
\label{sec:discussion_two_stage_training}

The two-stage training strategy progresses from
preserving local MLP functions to recovering overall
model performance.
Stage~I reconstructs each compressed MLP to approximately
preserve the input--output behavior of its teacher
counterpart, providing an initialization for Stage~II.
Stage~II jointly optimizes the compressed MLPs within
the full student model using ground-truth supervision
and teacher--student knowledge distillation, while
keeping all remaining student parameters frozen.
This separation is intended to reduce the initial
functional mismatch between the teacher and student
and mitigate the accumulation of errors across
compressed layers during end-to-end optimization.

\subsection{Influence of Teacher Capability and Training Data Distribution}
\label{sec:discussion_teacher_data}

The compressed student's performance also depends on
the teacher's capabilities and how closely the training
and downstream evaluation distributions match.
Poor teacher performance on a particular downstream
task may limit the usefulness of the distillation
supervision it provides for that task.
Moreover, a mismatch between the distributions of
the compression training data and the downstream
evaluation data may limit the transferability of
the compressed model.
Downstream results should therefore be interpreted
in the context of the teacher's baseline performance,
the training data distribution, and the selected
evaluation tasks.

\subsection{Limitations of the Proposed Framework}
\label{sec:discussion_limitations}

QuLoC has several limitations.
First, we have not benchmarked inference latency or
throughput against the uncompressed model or classical
compression baselines.
Although gain folding allows the model to run entirely
on conventional hardware without photonic-circuit
evaluation, this alone does not establish a speed advantage.

Second, the main model uses input-independent gain vectors.
This design allows the gains to be folded into the
low-rank factors after training, but the gains do not
adapt to individual tokens or hidden states.

Third, our experiments on Qwen3.5-4B and LLaMA-7B cover
only selected MLP layers and compression settings.
Further evaluation on other architectures and model
components is needed to determine how broadly the
method applies.
Additional runs with different random seeds and tests
under a wider range of photonic hardware conditions
are also needed to assess robustness.
\section{Conclusion}
\label{sec:conclusion}

We introduced QuLoC, a photonic quantum-assisted framework
for low-rank compression of MLPs in large language models.
It combines photonic gating of low-rank bottleneck
activations with a two-stage training strategy.
Stage~I trains each compressed MLP independently to
approximate the input--output behavior of its teacher
counterpart.
Stage~II then jointly refines the compressed MLPs within
the full student model using next-token supervision
and teacher--student knowledge distillation.
After training, the input-independent gain vectors
can be folded exactly into the low-rank factors,
allowing inference on conventional hardware without
photonic-circuit evaluation.

On Qwen3.5-4B, QuLoC reduces the language model's parameter
count by approximately $20.05\%$.
Among the compressed models evaluated, it achieves
the lowest perplexity on both WikiText-2 and C4 and
the highest average downstream accuracy.
The Stage~I ablation shows that local reconstruction
provides a better initialization for subsequent
end-to-end optimization.
QuLoC also performs competitively on LLaMA-7B.
The hardware evaluation further shows that measured
photonic features can support the proposed gating
mechanism.
Together, these results demonstrate that QuLoC can
recover a substantial portion of the original model's
performance on language modeling and downstream tasks
while using fewer parameters.

Future work will focus on refining QuLoC's training strategy
by better aligning training data with downstream
tasks.
We will evaluate these refinements across a broader range
of model architectures and conduct further experiments
on photonic quantum hardware to assess robustness under realistic
noise and measurement conditions.
\section{Acknowledgements}
\label{sec:ack}

This research is supported by the National Key R\&D Program of
China (Grant No.~2024YFA1409300 and 2024YFB4504005);
National Natural Science Foundation of China (NSFC)
(Grant Nos.~62235012, 12304342, 12574549, 12574542, and 125B1033);
Innovation Program for Quantum Science and Technology
(Grant Nos.~2021ZD0301500 and 2021ZD0300700);
Science and Technology Commission of Shanghai Municipality
(STCSM) (Grant Nos.~2019SHZDZX01, 24ZR1438700, 24ZR1430700,
and 24LZ1401500);
Startup Fund for Young Faculty at SJTU (SFYF at SJTU)
(Grant Nos.~24X010502876 and 24X010500170); Shanghai Municipal Commission of Economy and Informatization (no. 2025-GZL-RGZN-02047);
and Frontier Technologies R\&D Program of Jiangsu
(Grant No.~SBF20250000094).

\section{Description of generative AI contribution}
\label{sec:ai}
ChatGPT (OpenAI, GPT-6 Astra) was used to improve the language, clarity, and consistency of the manuscript, particularly in the methods, discussion, and conclusion sections. The authors independently developed the methodology, designed and conducted the experiments, and analyzed the results. Generative AI was not used to generate experimental data or results. All AI-assisted revisions were reviewed and approved by the authors, who take full responsibility for the final manuscript.


\bibliographystyle{unsrtnat}
\bibliography{refs}
\newpage
\appendix
\counterwithin{table}{section}
\counterwithin{figure}{section}
\counterwithin{algorithm}{section}
\section{Preliminaries and Problem Formulation}
\label{sec:preliminary}

\subsection{MLP Architectures of Qwen3.5-4B and Qwen3.5-27B}
\label{sec:qwen_mlp}

We use Qwen3.5-4B as the model to be compressed and additionally employ Qwen3.5-27B as an optional large teacher for end-to-end distillation.
Qwen3.5-4B contains 32 Transformer layers with a input dimension of $d=2560$ and a SwiGLU intermediate dimension of $m=9216$.
Qwen3.5-27B contains 64 Transformer layers with $d=5120$ and $m=17408$.

Both models employ SwiGLU MLPs. Given an input state $\mathbf{x}\in\mathbb{R}^{d}$, the MLP is computed as
\begin{equation}
\operatorname{MLP}(\mathbf{x})=W_{\mathrm{down}}\left[\operatorname{SiLU}\left(W_{\mathrm{gate}}\mathbf{x}\right)\odot W_{\mathrm{up}}\mathbf{x}\right],
\label{eq:qwen_swiglu}
\end{equation}
where
\begin{equation}
W_{\mathrm{gate}},W_{\mathrm{up}}\in\mathbb{R}^{m\times d},\qquad W_{\mathrm{down}}\in\mathbb{R}^{d\times m}.
\end{equation}

Ignoring bias terms, each SwiGLU MLP contains $3dm$ parameters.
Table~\ref{tab:qwen_mlp_config} summarizes the configurations and parameter counts of the two models.

\begin{table}[h]
\centering
\small
\caption{MLP configurations of the Qwen3.5 models used in this work.}
\label{tab:qwen_mlp_config}
\begin{tabular}{lrrrr}
\hline
Model & Layers & $d$ & $m$ & Params/MLP \\
\hline
Qwen3.5-4B  & 32 & 2560 & 9216  & 70.78M  \\
Qwen3.5-27B & 64 & 5120 & 17408 & 267.39M \\
\hline
\end{tabular}
\end{table}

The 32 MLP blocks in Qwen3.5-4B contain approximately $2.265$B parameters, whereas the 64 MLP blocks in Qwen3.5-27B contain approximately $17.113$B parameters.
In this work, only the MLPs of the 4B student are structurally compressed.
The 27B model remains uncompressed and is used only as an optional teacher in Stage~II.

\subsection{Photonic Quantum Circuits with 16 spatial modes}
\label{sec:photonic_preliminaries}

We construct the trainable photonic circuits using the path modes of a single photon.
For a system with $q$ optical modes, a single-photon state can be written as
\begin{equation}
\lvert\psi\rangle=\sum_{j=1}^{q}c_j\lvert j\rangle,\qquad\sum_{j=1}^{q}|c_j|^2=1,
\label{eq:single_photon_state}
\end{equation}
where $\lvert j\rangle$ denotes the photon occupying the $j$-th path and $c_j$ is the corresponding complex amplitude.
A lossless linear-optical circuit composed of beam splitters and phase shifters applies a unitary transformation to this state.

The Mach--Zehnder interferometer (MZI) is the basic trainable unit of the circuit.
An MZI consists of two balanced beam splitters and tunable phase shifters and implements a $2\times2$ unitary transformation over two adjacent modes:
\begin{equation}
U_{\mathrm{MZI}}(\theta,\phi)=B R(\theta) B R(\phi),
\label{eq:mzi}
\end{equation}
where
\begin{equation}
B=\frac{1}{\sqrt{2}}
\begin{bmatrix}
1 & i\\
i & 1
\end{bmatrix},
\qquad R(\varphi)=\begin{bmatrix}
1 & 0\\
0 & e^{i\varphi}
\end{bmatrix}.
\end{equation}
A Clements interferometer arranges multiple MZIs in a rectangular mesh to implement a multimode unitary transformation.
Denoting all trainable phase parameters by $\boldsymbol{\theta}$, the circuit output is
\begin{equation}
\lvert\psi_{\mathrm{out}}\rangle=U(\boldsymbol{\theta})\lvert\psi_{\mathrm{in}}\rangle.
\label{eq:photonic_unitary}
\end{equation}

Single-photon detection at the output modes produces the probability features
\begin{equation}
z_j(\boldsymbol{\theta})=\left|\langle j\mid\psi_{\mathrm{out}}\rangle\right|^2,\qquad j=1,\ldots,q.
\label{eq:photonic_probability}
\end{equation}
The resulting vector satisfies
\begin{equation}
\mathbf{z}(\boldsymbol{\theta})=[z_1,\ldots,z_q]^\top,\qquad\mathbf{z}\geq\mathbf{0},\qquad\mathbf{1}^{\top}\mathbf{z}=1.
\label{eq:photonic_feature}
\end{equation}
We use $q=16$, so each circuit produces a 16-dimensional measurement-probability vector $\mathbf{z}(\boldsymbol{\theta})$.
This vector is differentiable with respect to the circuit phases and is subsequently used to generate the low-rank channel gains.
\section{Compression Configuration Ablations}
\label{app:compression_configuration_ablation}

This appendix examines the effects of compression configurations
and evaluates the performance gains from photonic gating.
Appendix~\ref{app:rank_ablation} investigates the effect of the
low-rank dimension under a fixed compression scope, whereas
Appendix~\ref{app:depth_ablation} studies the effect of the number
of compressed MLP layers under central-layer placement.
These two analyses characterize the compression--performance
trade-off and inform the configuration adopted in the main experiments.
Separately, Appendix~\ref{app:gating_depth} evaluates how the
performance gains from gating vary with the number of compressed
layers.

\subsection{Effect of Rank on the Trade-off between Compression and Performance}
\label{app:rank_ablation}

We investigate the effect of the low-rank dimension on the trade-off between compression and performance of the proposed method. To isolate the
effect of rank, the compression scope and training configuration are kept
fixed, and only the rank \(r\) is varied. Specifically, the middle four MLP
modules in Qwen3.5-4B are compressed with \(r\in\{256,384,512\}\).
This experiment is intended to characterize the effect of rank under a fixed
compression scope rather than to achieve the target model-level compression
ratio used in the main experiments. The results are reported in
Table~\ref{tab:rank_ablation}.

\begin{table}[t]
    \centering
    \scriptsize
    \renewcommand{\arraystretch}{1.25}
    \setlength{\tabcolsep}{3.5pt}
    \caption{
    Rank ablation on the Qwen3.5-4B model.
    The middle four MLP layers are compressed using the proposed method.
    The column ``Total Parameters after Compression'' reports the total number
    of parameters in the resulting compressed model.
    The parameter reduction rate denotes the proportion of parameters removed
    relative to the original model.
    Parameter counts are reported to $0.001\,\mathrm{B}$; reductions and ratios are calculated from unrounded counts and reported to two decimal places.
    For reference, the original uncompressed Qwen3.5-4B model achieves a
    WikiText-2 PPL of $9.570$ under the same evaluation protocol.
    Lower WikiText-2 perplexity (PPL) is better.
    }
    \label{tab:rank_ablation}
    \begin{tabular}{ccccc}
        \toprule
        Rank
        & Total Parameters after Compression
        & Parameter Reduction Rate (\%)
        & Model Compression Ratio ($\times$)
        & WikiText-2 PPL $\downarrow$ \\
        \midrule
        256 & $3.971\,\mathrm{B}$ & $5.58$ & $1.06$ & $10.170$ \\
        384 & $3.995\,\mathrm{B}$ & $5.01$ & $1.05$ & $10.142$ \\
        512 & $4.019\,\mathrm{B}$ & $4.44$ & $1.05$ & $10.068$ \\
        \bottomrule
    \end{tabular}
\end{table}

As the rank increases from \(256\) to \(512\), the total number of parameters
in the compressed model increases from \(3.971\)B to \(4.019\)B, while the
parameter reduction rate decreases from \(5.58\%\) to \(4.44\%\). Each
increase of 128 in rank introduces approximately \(0.024\)B additional
parameters. This behavior follows directly from the low-rank
parameterization: increasing \(r\) introduces more rank channels and therefore
more trainable degrees of freedom, at the cost of a lower compression rate.

Meanwhile, language modeling performance improves consistently with increasing
rank. The WikiText-2 PPL decreases from \(10.170\) at \(r=256\) to
\(10.142\) at \(r=384\), and further to \(10.068\) at \(r=512\).
Compared with \(r=256\), increasing the rank to \(384\) and \(512\)
reduces PPL by \(0.028\) and \(0.102\), respectively. This trend shows that a larger rank
alleviates the information bottleneck imposed by low-rank compression and
allows the compressed MLPs to preserve more of the original model
functionality. However, this improvement is obtained at the expense of a
progressively lower parameter reduction rate.

The three settings therefore represent different operating points along the
compression--performance trade-off. Rank \(256\) provides the highest
parameter reduction but also the largest PPL degradation, whereas rank \(512\)
achieves the best language modeling performance with the lowest compression
rate among the evaluated settings. Rank \(384\) lies between these two
extremes, retaining a parameter reduction rate of approximately \(5.01\%\)
while reducing the PPL from \(10.170\) to \(10.142\).
We therefore use
\(r=384\) as a balanced operating point in the main comparative experiments, unless
otherwise specified.

\subsection{Effect of the Number of Compressed Layers}
\label{app:depth_ablation}

We further investigate how the number of compressed MLP layers affects the model-level compression rate and language modeling performance.
All experiments are conducted on Qwen3.5-4B, where a contiguous block of centrally located MLP modules is replaced by the proposed photonic quantum-gated low-rank modules.
Two ranks, $r=384$ and $r=512$, are evaluated under the same compression placement strategy.
The results are reported in Table~\ref{tab:compression_depth_ablation}.

At a fixed rank, increasing the number of compressed MLP layers monotonically increases the overall parameter reduction.
For example, at $r=384$, the parameter reduction rate increases from $5.01\%$ for four layers to $10.02\%$, $20.05\%$, $30.07\%$, and $40.09\%$ for 8, 16, 24, and all 32 layers, respectively.
A similar trend is observed for $r=512$.
This behavior is expected because expanding the compression scope replaces a larger fraction of the original MLP parameters with low-rank representations.

The effect on language modeling performance becomes increasingly pronounced as more layers are compressed.
At $r=384$, the WikiText-2 PPL increases only slightly from 10.142 with four compressed layers to 10.319 with eight layers.
It then rises to 12.173 and 16.161 when 16 and 24 layers are compressed, respectively. Compressing all 32 MLP layers causes the PPL to increase sharply to 1904.825, indicating a severe degradation of language modeling capability.
A similar trend is observed at $r=512$, where the PPL changes from 10.068 and 10.322 for four and eight compressed layers to 11.974 and 14.677 for 16 and 24 layers, before increasing to 1223.689
when all 32 layers are compressed. These results suggest that the model can tolerate the approximation errors introduced by a moderate increase in the number of compressed MLP layers, whereas compressing nearly all MLP layers substantially degrades model performance.

Increasing the rank generally mitigates the degradation caused by deeper compression, although the difference is negligible when only a small number of layers are compressed.
For four compressed layers, increasing the rank from 384 to 512 reduces PPL from 10.142 to 10.068, while the two settings are effectively tied for eight layers (10.319 versus 10.322).
The benefit becomes more visible at larger compression depths: the PPL decreases from 12.173 to 11.974 for 16 layers and from 16.161 to 14.677 for 24 layers.
When all 32 layers are compressed, both rank settings suffer severe performance degradation despite the larger rank.

Overall, these results reveal a clear trade-off between compression depth, retained rank, and language modeling performance.
Increasing the number of compressed layers yields greater model-level parameter reduction, but the associated performance degradation becomes substantially more severe under aggressive compression.
The intermediate configuration with $r=384$ and 16 compressed layers provides a 20.05\% parameter reduction while maintaining a WikiText-2 PPL of 12.173, and is therefore adopted in the main Qwen3.5-4B comparison.

\begin{table}[t]
    \centering
    \renewcommand{\arraystretch}{1.25}
    \scriptsize
    \setlength{\tabcolsep}{3.5pt}
    \caption{Effect of the number of compressed MLP layers under
    central-layer placement on Qwen3.5-4B. Parameter counts and
    reductions refer to the language-model component, excluding the
    vision encoder. The compression ratio is the original parameter
    count divided by the compressed parameter count. Parameter counts
    are rounded to $0.001\,\mathrm{B}$; reductions and ratios are
    calculated from unrounded parameter counts and reported to two decimal places. The uncompressed model achieves a
    WikiText-2 PPL of 9.570. Lower PPL is better.}
    \label{tab:compression_depth_ablation}
    \resizebox{\linewidth}{!}{%
    \begin{tabular}{l|c|cc|c|c}
        \toprule
        \textbf{Compression Location}
        & \textbf{Rank}
        & \textbf{Total Params.}
        & \textbf{Reduction (\%)}
        & \textbf{Ratio ($\times$)}
        & \textbf{WikiText-2 PPL} $\downarrow$ \\
        \midrule
        Middle 4 Layers
        & 384 & $3.995\,\mathrm{B}$ & 5.01 & 1.05 & 10.142 \\
        & 512 & $4.019\,\mathrm{B}$ & 4.44 & 1.05 & 10.068 \\
        \midrule
        Middle 8 Layers
        & 384 & $3.784\,\mathrm{B}$ & 10.02 & 1.11 & 10.319 \\
        & 512 & $3.832\,\mathrm{B}$ & 8.88 & 1.10 & 10.322 \\
        \midrule
        Middle 16 Layers
        & 384 & $3.363\,\mathrm{B}$ & 20.05 & 1.25 & 12.173 \\
        & 512 & $3.459\,\mathrm{B}$ & 17.75 & 1.22 & 11.974 \\
        \midrule
        Middle 24 Layers
        & 384 & $2.941\,\mathrm{B}$ & 30.07 & 1.43 & 16.161 \\
        & 512 & $3.086\,\mathrm{B}$ & 26.63 & 1.36 & 14.677 \\
        \midrule
        All 32 Layers
        & 384 & $2.520\,\mathrm{B}$ & 40.09 & 1.67 & 1904.825 \\
        & 512 & $2.713\,\mathrm{B}$ & 35.50 & 1.55 & 1223.689 \\
        \bottomrule
    \end{tabular}%
    }
\end{table}

\subsection{Gating Gains across Different Numbers of Compressed Layers}
\label{app:gating_depth}

We investigate how the gains from photonic gating vary
with the number of compressed layers. We fix the low-rank
dimension at $r=512$ and compress the MLP modules in the last
4, 8, or 12 transformer layers. Our method is compared with
\textit{Optimized SVD}, using the same
truncated-SVD initialization for $P$ and $B$ and the same
two-stage training procedure.
Table~\ref{tab:gating_depth} reports normalized accuracy on
PIQA, ARC-Easy, and OpenBookQA, along with the average accuracy
across the three tasks (Avg.).
Each method's accuracy row is followed by its absolute and
relative drops from the uncompressed teacher.
For teacher accuracy $A_{\mathrm{T}}$ and compressed-model
accuracy $A_{\mathrm{C}}$, the absolute drop is
$100(A_{\mathrm{T}}-A_{\mathrm{C}})$ in percentage points,
and the relative drop is
$(A_{\mathrm{T}}-A_{\mathrm{C}})/A_{\mathrm{T}}\times100\%$.
Smaller drops indicate better preservation of teacher performance.
For Avg., the absolute and relative drops are computed from the teacher's and compressed model's three-task mean accuracies using the same formulas.

The results in Table~\ref{tab:gating_depth} show that both methods lose more average accuracy relative to the teacher
as the number of compressed layers increases, but quantum
gating consistently limits this degradation and provides
larger benefits when more layers are compressed. Specifically, quantum gating reduces the loss in average
accuracy relative to the teacher by 0.27, 5.70, and 8.07
percentage points compared with the ungated baseline
when the last 4, 8, and 12 layers are compressed, respectively. One possible explanation is that gating allows the model
to adjust the contribution of each low-rank component
during training, which may help reduce errors introduced
by compression. In summary, these results demonstrate the effectiveness of quantum gating
and show that its gains increase with the number of compressed
layers when both methods use the same initialization of $P$
and $B$ and the same two-stage training procedure.

\begin{table*}[t]
\centering
\caption{Downstream accuracy (reported as fractions), absolute drops (pp), and relative drops (\%) from the uncompressed teacher at $r=512$. Avg.\ denotes the three-task mean. Average drops are calculated before rounding the mean accuracy.}
\label{tab:gating_depth}
\small
\setlength{\tabcolsep}{6pt}
\renewcommand{\arraystretch}{1.2}
\begin{tabular}{llrrrr}
\toprule
Compressed layers & Method / Metric
& PIQA & ARC-Easy & OpenBookQA & Avg. \\
\midrule
 & Teacher
& 0.7851 & 0.7551 & 0.4060 & 0.6487 \\
\midrule

Last 4 & \textit{Optimized SVD}
& 0.7639 & 0.7269 & 0.4020 & 0.6309 \\
& \quad Absolute drop (pp)
& 2.12 & 2.82 & 0.40 & 1.78 \\
& \quad Relative drop (\%)
& 2.70 & 3.73 & 0.99 & 2.74 \\
\addlinespace
& QuLoC
& \textbf{0.7671} & \textbf{0.7277}
& \textbf{0.4060} & \textbf{0.6336} \\
& \quad Absolute drop (pp)
& 1.80 & 2.74 & 0.00 & 1.51 \\
& \quad Relative drop (\%)
& 2.29 & 3.63 & 0.00 & 2.33 \\
\midrule

Last 8 & \textit{Optimized SVD}
& 0.6920 & 0.6090 & 0.3600 & 0.5537 \\
& \quad Absolute drop (pp)
& 9.31 & 14.61 & 4.60 & 9.51 \\
& \quad Relative drop (\%)
& 11.86 & 19.35 & 11.33 & 14.65 \\
\addlinespace
& QuLoC
& \textbf{0.7560} & \textbf{0.6940}
& \textbf{0.3820} & \textbf{0.6107} \\
& \quad Absolute drop (pp)
& 2.91 & 6.11 & 2.40 & 3.81 \\
& \quad Relative drop (\%)
& 3.71 & 8.09 & 5.91 & 5.87 \\
\midrule

Last 12 & \textit{Optimized SVD}
& 0.6620 & 0.5620 & 0.3400 & 0.5213 \\
& \quad Absolute drop (pp)
& 12.31 & 19.31 & 6.60 & 12.74 \\
& \quad Relative drop (\%)
& 15.68 & 25.57 & 16.26 & 19.64 \\
\addlinespace
& QuLoC
& \textbf{0.7550} & \textbf{0.6630}
& \textbf{0.3880} & \textbf{0.6020} \\
& \quad Absolute drop (pp)
& 3.01 & 9.21 & 1.80 & 4.67 \\
& \quad Relative drop (\%)
& 3.83 & 12.20 & 4.43 & 7.20 \\
\bottomrule
\end{tabular}
\end{table*}
\end{document}